\documentclass[sigconf, authorversion]{acmart}
\AtBeginDocument{%
  }

\copyrightyear{2026}
\acmYear{2026}
\setcopyright{cc}
\setcctype{by}
\acmConference[ICER 2026 Vol. 1]{Proceedings of the ACM Conference on International Computing Education Research Vol.1}{August 11--14, 2026}{Uppsala, Sweden}
\acmBooktitle{Proceedings of the ACM Conference on International Computing Education Research Vol.1 (ICER 2026 Vol. 1), August 11--14, 2026, Uppsala, Sweden}
\acmDOI{10.1145/3765964.3811667}
\acmISBN{979-8-4007-2203-5/2026/08}

\usepackage[utf8]{inputenc}
\usepackage{multirow}
\usepackage{graphicx}
\usepackage{color}
\usepackage{subcaption}
\usepackage[inline]{enumitem}
\usepackage{makecell}

\definecolor{diffold}{rgb}{0.82,0.93,0.82}
\definecolor{diffnew}{rgb}{0.988,0.922,0.922}

\newcommand{\oldline}[1]{%
  \begingroup
  \setlength{\fboxsep}{0.6pt}%
  \colorbox{diffold}{\texttt{#1}}%
  \endgroup
}

\newcommand{\bugline}[1]{%
  \begingroup
  \setlength{\fboxsep}{0.6pt}%
  \colorbox{diffnew}{\texttt{#1}}%
  \endgroup
}

\begin{document}


\title[When AI Is Wrong on Purpose: How Students Respond to Buggy GenAI Code]{When AI Is Wrong on Purpose: \\ How Students Respond to Buggy GenAI Code}

\author{Victor-Alexandru P{\u a}durean}
\affiliation{%
  \institution{MPI-SWS}
  \city{Saarbr{\"u}cken}
  \country{Germany}
}
\email{vpadurea@mpi-sws.org}
\orcid{0009-0004-2998-096X}

\author{Kaitlin Riegel}
\affiliation{%
  \institution{University of Auckland}
  \city{Auckland}
  \country{New Zealand}
}
\email{kaitlin.riegel@auckland.ac.nz}
\orcid{0000-0002-8187-2016}

\author{Alkis Gotovos}
\affiliation{%
  \institution{MPI-SWS}
  \city{Saarbr{\"u}cken}
  \country{Germany}
}
\email{agkotovo@mpi-sws.org}
\orcid{0000-0002-3902-8890}

\author{Jyotika Mahapatra}
\affiliation{%
  \institution{MPI-SWS}
  \city{Saarbr{\"u}cken}
  \country{Germany}
}
\email{jmahaptra@mpi-sws.org}
\orcid{0009-0004-0551-231X}

\author{Ahana Ghosh}
\affiliation{%
  \institution{MPI-SWS}
  \city{Saarbr{\"u}cken}
  \country{Germany}
}
\email{gahana@mpi-sws.org}
\orcid{0000-0002-0967-5886}

\author{Paul Denny}
\affiliation{%
  \institution{University of Auckland}
  \city{Auckland}
  \country{New Zealand}
}
\email{paul@cs.auckland.ac.nz}
\orcid{0000-0002-5150-9806}

\author{Juho Leinonen}
\affiliation{%
  \institution{Aalto University}
  \city{Espoo}
  \country{Finland}
}
\email{juho.2.leinonen@aalto.fi}
\orcid{0000-0001-6829-9449}

\author{James Prather}
\affiliation{%
  \institution{Abilene Christian University}
  \city{Abilene}
  \state{TX}
  \country{USA}
}
\email{james.prather@acu.edu}
\orcid{0000-0003-2807-6042}

\author{Adish Singla}
\affiliation{%
  \institution{MPI-SWS}
  \city{Saarbr{\"u}cken}
  \country{Germany}
}
\email{adishs@mpi-sws.org}
\orcid{0000-0001-9922-0668}

\renewcommand{\shortauthors}{Victor-Alexandru P{\u a}durean et al.}

\begin{abstract}
As Generative AI (GenAI) becomes increasingly central to software development, CS education is integrating prompt-centered workflows where students describe intended program behavior in natural language to elicit code. However, professional practice requires careful review and verification of GenAI-generated code that may appear correct while containing subtle faults. This creates a challenge for CS1-level activities, where current models often solve tasks correctly and reduce students' incentive to closely inspect generated outputs. We investigate how prompt-centered programming activities can be adapted to better foster these practices. Specifically, we explore an approach where realistic, runnable bugs are injected into otherwise correct solutions, thus requiring students to read and repair generated outputs. We analyzed $2{,}636$ sessions from $917$ students, and examined behavior across instances of naturally occurring prompt-related failures and deliberately injected bugs within each session. Our findings show that students responded differently across bug sources. Deliberately injected bugs more often led to direct code edits and higher next-attempt success, suggesting localized repair of near-miss solutions. Prompt-related failures instead more often led students to refine prompts by clarifying constraints, updating function signatures, adding edge cases, or reframing the task. Student reflections reinforce the emphasis on review and repair, describing useful practice in code understanding, code review, and debugging, as well as a more careful verification mindset and greater awareness of GenAI limitations. Ultimately, prompt-related failures and injected bugs together support a pedagogically useful GenAI workflow, where students practice both specification refinement through prompts and debugging through code editing.
\end{abstract}



\maketitle  

\section{Introduction}
\label{sec:introduction}

\begin{figure*}[t!]
\centering
	\includegraphics[width=0.95\textwidth,trim={0 1mm 27mm 1mm},clip]{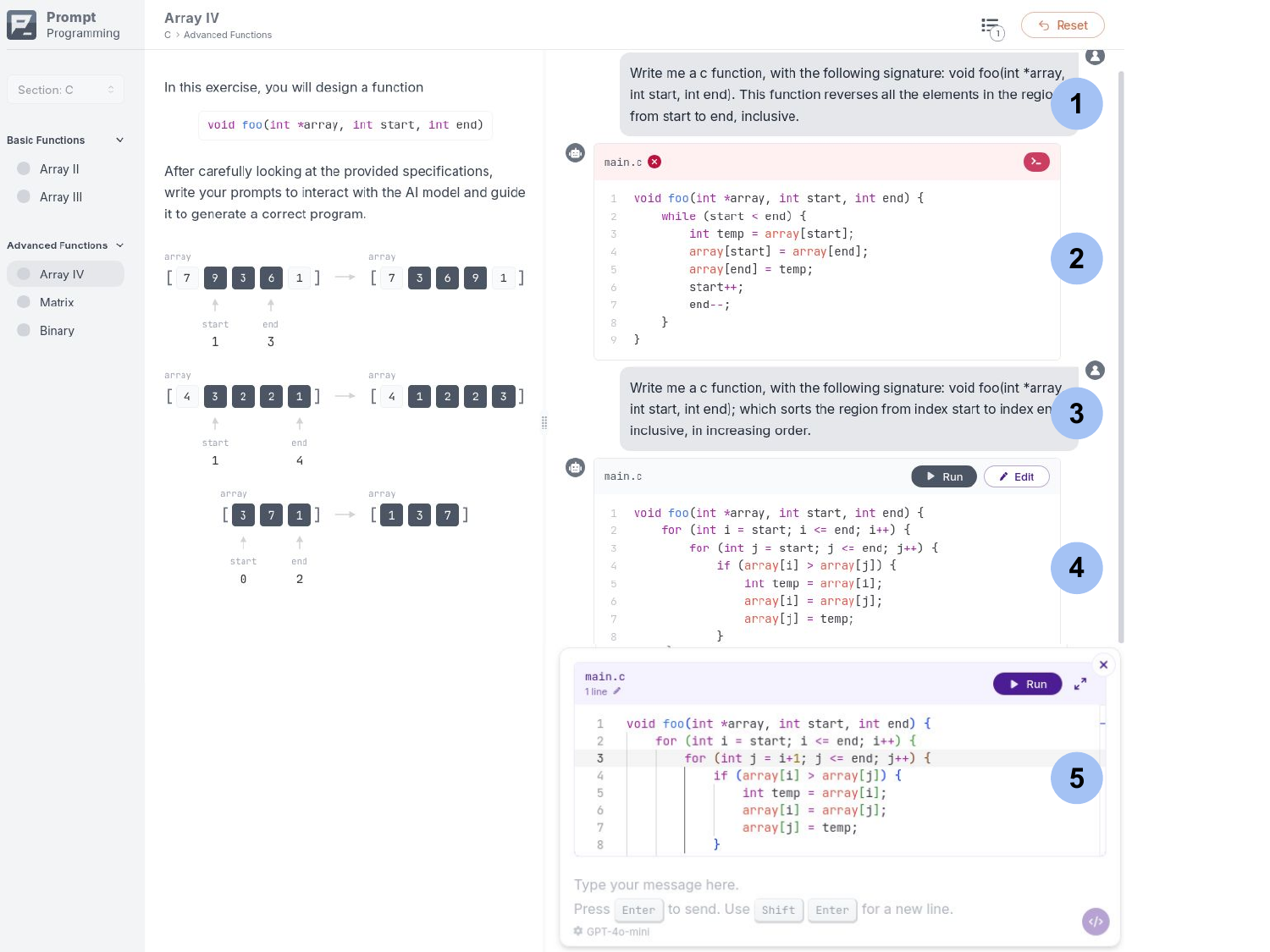}
    \caption{\looseness-1The interface of Prompt Programming \cite{padurean2025prompt}, exemplified on Array IV. The interaction proceeds as follows: (1) the student sends an initial prompt that mis-specifies the task (reversing the array segment from \texttt{start} to \texttt{end}); (2) the GenAI assistant responds with a reversing implementation, which is incorrect for the intended sorting task; (3) the student reprompts, correcting the request to sort the segment in ascending order; (4) the GenAI assistant generates a correct sorting solution, but our bug injection pipeline intercepts it and shows a near-miss version where the inner loop starts \texttt{j} at \texttt{start} instead of \texttt{i+1}; and (5) the student edits the code manually in the editor.}
    \Description[The interface for our tool, showing chat, code, and an edit]{Screenshot of our tool's web interface. A left sidebar lists course tasks and shows the Array IV task statement with array diagrams. The main panel shows a chat transcript with five numbered markers placed next to successive messages. Assistant messages include C code blocks. The last step shows the built-in code editor containing the assistant's code, with a visible cursor area and a Run button below, indicating the student is making a manual edit after receiving a failing solution.}
    \label{fig.illustration}
\end{figure*}

\looseness-1Widespread adoption of Generative AI (GenAI) coding assistants is transforming professional software development. Most developers now integrate AI tools into their workflows \cite{stackoverflow,DBLP:journals/cacm/ZieglerKLRRSSA24}, and recent work has begun to explore how these tools redefine traditional development roles and practices \cite{DBLP:journals/ase/SauvolaTKRD24,DBLP:journals/infsof/BanhHS25}. However, the ability to generate code quickly does not guarantee correctness, especially at higher levels of complexity in large code bases. Empirical evaluations of copilot-style suggestions show substantial rates of incorrect solutions even on well-scoped programming tasks \cite{mo2025assessing}. Studies of GenAI-generated code report recurring bug patterns \cite{tambon2025bugs} and frequent security vulnerabilities \cite{DBLP:journals/ese/TihanyiBFJC25}. Output variability also means developers often need to inspect and compare multiple alternatives rather than relying on a single suggestion \cite{DBLP:journals/infsof/OertelKH25}. Consequently, effective use of GenAI requires skills in clear specification as well as critical code review, systematic testing, and debugging. These needs are reflected in emerging workflows such as GenAI-supported code review that explicitly leverage prompt engineering and refinement \cite{DBLP:journals/infsof/PornprasitT24}. Computing education has mirrored these developments, integrating GenAI-based workflows into the curriculum. Educators have called for curricular adaptation \cite{DBLP:journals/cacm/DennyPBFHLLRSS24} and the community is actively debating implications for learning goals and assessment \cite{DBLP:conf/sigcse/MacNeil00KHPBWR24}. Early CS1 course designs integrating LLMs emphasize fundamentals such as decomposition, explanation, and testing when producing software with GenAI \cite{DBLP:conf/iticse/VadapartyZSPAB024}. Broader studies further argue for treating GenAI use as a first-class curricular concern, framing risks and opportunities while outlining concrete teaching responses \cite{DBLP:conf/iticse/Prather00BACKKK23,DBLP:journals/corr/abs-2402-01580}.

\looseness-1A growing line of work has explored how to incorporate GenAI into computing education through activities centered on \emph{natural-language specification}. Prompt Problems, prompt-based programming, and related instructional designs treat prompting as a central learning goal, examining how students formulate and refine prompts to communicate intent \cite{DBLP:conf/sigcse/00010PLABR24,kerslake2024integrating,padurean2025prompt}. However, this emphasis can shift attention away from inspecting generated code. Empirical studies report that students frequently submit AI-generated solutions with minimal modification \cite{kazemitabaar2023studying}, and that novices often rely on models to produce and iteratively fix solutions without critically reasoning about the code \cite{DBLP:journals/pacmse/RaheM25}. More broadly, recent work shows that students may struggle to comprehend GenAI code while simultaneously developing an illusion of competence, believing they understand solutions they have not critically evaluated \cite{DBLP:conf/sigsoft/ZiLGAF25,prather2024widening}. These patterns are reinforced by the fact that, for many CS1-level tasks, modern models often produce correct solutions. \emph{If the generated code always seems to work, what incentive do students have to inspect it closely?} This stands in contrast to professional practice, where code that appears correct may contain subtle faults and must be carefully reviewed, tested, and debugged. Within GenAI-mediated workflows at the introductory level, students may have few structured opportunities to practice verification and repair.

\looseness-1To address this gap, we explore whether deliberately introducing small, realistic faults into otherwise correct GenAI-generated code can create valuable code review opportunities for students. To do this, we integrate a bug injection pipeline into Prompt Programming \cite{padurean2025prompt}, our publicly available prompt-based programming platform. Students receive a task and write a natural language prompt which a GenAI assistant uses to generate code which is then run against hidden tests. If unsuccessful, they can iteratively respond to failures by engaging  with the assistant or the code. We distinguish two scenarios.  In the first case, the student-written prompt generates code that does not pass all the tests (we call this a \emph{`natural bug'}) and we show the test case failures along with the code. In the second case, the student-written prompt generates code that \emph{would} pass the tests. In this latter case, our pipeline replaces that code with a realistic, runnable near-miss variant containing a subtle fault (we call this an \emph{`injected bug'}). In either case, the student can try to resolve the bug either by fixing the code directly or by modifying the prompt they send to the GenAI model (which will subsequently result in another natural or injected bug).

\looseness-1We analyze student behaviors with this pipeline at scale in a classroom deployment, covering $2{,}636$ student \emph{sessions} (a complete sequence of prompts and code edits for a given task that ends when the student completes or abandons the task) and $6{,}071$ bug-fixing \emph{turns} (individual student follow-ups after a GenAI response with buggy code, natural or injected). Figure~\ref{fig.illustration} illustrates the platform interface and a representative session in which a natural bug is followed by an injected bug, highlighting a workflow that requires both precise specification through prompting and careful code review to resolve.

Our goal with this work is not to undermine students' trust in GenAI or their own abilities, but to make code review an explicit part of  using GenAI. In our design, injected bugs are small, realistic, and accompanied by feedback that mirrors authentic development workflows in which externally authored code, including AI-generated code, must be reviewed, tested, and repaired. This design creates structured opportunities for students to inspect generated code, verify its behavior, and make localized repairs, instead of relying only on passive acceptance of generated solutions.

We leverage large-scale classroom data to evaluate how students interact with this integrated environment, focusing on behavioral patterns and qualitative feedback. Specifically, our analysis is guided by the following research questions:

\begin{enumerate}[label=\textbf{RQ\arabic*:},leftmargin=3em]
    \item How do different bug sources (natural vs. injected) influence students' immediate response choices and immediate outcomes when interacting with buggy GenAI code?

    \item How do students' repair strategies differ across bug sources after encountering buggy GenAI code?

    \item How do students reflect on their experience, understanding of AI-generated code, and perceived learning benefits of prompting tasks with deliberately buggy GenAI code?

\end{enumerate}

\looseness-1The rest of this paper is organized as follows. Section~\ref{sec:related-work} reviews related work, Section~\ref{sec:method} describes the study methodology, Section~\ref{sec:results} presents the results, Section~\ref{sec:discussion} discusses implications and limitations, while Section~\ref{sec:conclusion} concludes our paper.



\section{Related Work}
\label{sec:related-work}

In this section, we discuss several areas of related work that motivate our study and help position our contributions.

\subsection{Generative AI and Agents in Computing Education}

\looseness-1GenAI, especially code-generating large language models (LLMs), has rapidly become central in computing education. Position papers and broad reviews argue that on-demand code and explanations change what it means to learn programming, and push the community to revisit pedagogy, assessment, and learning goals in light of widespread access to these tools \cite{DBLP:conf/iticse/Prather00BACKKK23,DBLP:journals/corr/abs-2402-01580,DBLP:journals/cacm/DennyPBFHLLRSS24,prather2024beyond}. Recent work also reflects the expanding capabilities of these systems, including multimodal interactions in programming support and more agentic workflows in which LLMs take on multiple roles, such as generation and validation \cite{Hou2026exploring,DBLP:conf/iticse/OuhTLG25,DBLP:conf/aied/NguyenPGTS25,DBLP:conf/lak/PhungPS0CGSS24}.
Building on the potential of GenAI, CS education research has been studying concrete applications that help students during programming. Some examples include improved error messages, targeted feedback for syntax errors, providing code repairs, and semantic hint generation \cite{DBLP:conf/sigcse/0001HSRDPB23,DBLP:conf/edm/PhungCGKMSS23,DBLP:conf/sigcse/WangMP24,DBLP:journals/pacmpl/ZhangCGLPSV24,DBLP:conf/lak/PhungPS0CGSS24,DBLP:conf/nips/KotalwarGS24}. Other studies examine GenAI as an interactive programming partner, including copilot-style support and chat-based assistants for multi-turn problem solving \cite{DBLP:journals/tochi/PratherRDBLLPFS24,DBLP:conf/aied/VibergWFDE25,DBLP:conf/iticse/SchollK25,padurean2025prompt}. In parallel, instructor-facing work studies how to integrate GenAI into classrooms while preserving pedagogical control, for example through assistants that avoid giving away full solutions and emphasize transparency and guardrails~\cite{kazemitabaar2024codeaid,DBLP:conf/iticse/Prather00BACKKK23,prather2024beyond}. It also includes systems that generate programming learning materials, such as exercises, through automatic generation, personalization, and task synthesis~\cite{DBLP:conf/icer/SarsaDH022,DBLP:conf/icer/LogachevaHPS024,DBLP:conf/aied/NguyenPGTS25}.

A recurring concern in this rapid incorporation of GenAI across computing education is overreliance. GenAI can accelerate students who already have a plan, but it can also amplify metacognitive difficulties, create illusions of competence, and reduce engagement with code reasoning and verification \cite{prather2024widening,kazemitabaar2024codeaid}. As a result, several CS education papers emphasize designing interventions that keep students actively involved (e.g., requiring checking and justification) and explicitly preparing learners to be critical, informed GenAI users \cite{Chandrashekar2026demistify,DBLP:conf/iticse/Prather00BACKKK23}. This sets up the need to study not just whether GenAI helps, but how learners respond when GenAI outputs fail, and which structured failure experiences support debugging and code understanding without outsourcing them to the model.

\subsection{Prompt Problems and Prompt-Based Programming}

\looseness-1Recent work increasingly treats prompts designed for GenAI to produce code as a first-class programming artifact rather than a simple request for help. Prompt-centered exercises make natural-language specification and refinement central to the task. Specifically, Prompt Problems operationalize these skills by assessing students on their ability to prompt an AI model to generate code that meets a target specification \cite{DBLP:conf/sigcse/00010PLABR24}. Recent work has shown how prompting-for-code can be integrated into introductory programming through structured labs that emphasize communicating intent and evaluating whether generated code meets task requirements \cite{kerslake2024integrating}. Beyond one-shot exercises, prompt-based programming is increasingly framed as an interactive workflow where students engage in multi-turn dialogue, run generated code, interpret failures, and revise their requests \cite{padurean2025prompt,DBLP:conf/sigcse/YehTGYFC25}. This further aligns with broader arguments that natural-language prompting is becoming a significant abstraction layer in computing education \cite{DBLP:conf/kolicalling/ReevesP00MLNB25}. Together, this work shifts attention from one-shot generation to the prompt–test–revise loop students follow when outputs are plausible but incorrect. In this study, we build on these foundations to investigate how learners respond to targeted failures, specifically exploring how we can encourage attentive code review and teach students to navigate the inherent dangers of overreliance on GenAI \cite{DBLP:journals/tochi/PratherRDBLLPFS24,prather2024widening,DBLP:conf/kolicalling/KazemitabaarHHE23}. Using natural and injected bugs to create distinct failure contexts, we examine how near-miss failures can serve as a teaching tool to promote code inspection, verification, and debugging rather than passive acceptance of AI-generated code.

\subsection{Debugging, Bug Injection, and Controlled Failures}

Debugging is consistently reported as one of the hardest parts of learning to program, largely because novices struggle with the steps between seeing a failure and producing a concrete, testable explanation. Prior work shows beginners often follow an edit-and-test approach, making superficial code changes without a clear hypothesis, instead of systematically localizing the fault and validating hypotheses with evidence \cite{mccauley2008debugging,fitzgerald2008debugging}. Such patterns show up across multi-institution studies of novice debuggers and qualitative analyses of their tactics (e.g., tracing, print debugging, commenting out code, or small experimental edits) \cite{DBLP:conf/sigcse/MurphyLMSTZ08,DBLP:journals/hhci/KatzA87}. These findings motivate explicit, structured debugging instruction, since debugging skill does not automatically follow from code-writing skill \cite{DBLP:conf/wipsce/MichaeliR19}. The need persists in GenAI-supported workflows, as even with an AI debugging tutor, novices vary in when they seek help and how they engage with suggestions \cite{DBLP:conf/icer/YangZXBS24}. Interviews suggest they do not treat the AI tutor as a primary source for learning debugging strategies. A recent systematic review also shows wide variation in how and how much debugging is taught across curricula \cite{DBLP:journals/toce/YangBOBS24}.

A common instructional response is to use controlled failures, including seeded bugs and injected faults, so students can practice connecting symptoms to causes \cite{mccauley2008debugging,fitzgerald2008debugging,DBLP:conf/wipsce/MichaeliR19}. GenAI systems also make controlled-failure practice easier to scale, with tools like BugSpotter generating validated buggy programs for debugging practice \cite{DBLP:conf/sigcse/Padurean0S25}, while HypoCompass uses a learn-by-teaching setup that pushes students to articulate and test hypotheses during debugging \cite{DBLP:conf/aied/MaSKW24}. Related debugging-tool research also emphasizes explanation, for example by supporting explicit ``why'' and ``why not'' questions about program behavior \cite{DBLP:conf/icse/KoM08}. However, this line of work largely treats debugging practice as separate from end-to-end GenAI programming workflows, and does not study deliberately injecting realistic faults into otherwise plausible model-generated solutions during prompt-based development.



\section{Methodology}\label{sec:method}
\looseness-1In this section, we describe the study setting and analysis approach. We first introduce the Prompt Programming platform, then detail our bug injection pipeline, and summarize classroom deployment, including the five lab tasks. Next, we define key concepts, data collection, and describe the analysis procedures.

\subsection{The Prompt Programming Platform}\label{sec:method.platform}

\begin{figure*}[t!]
\centering
	\includegraphics[width=0.98\textwidth]{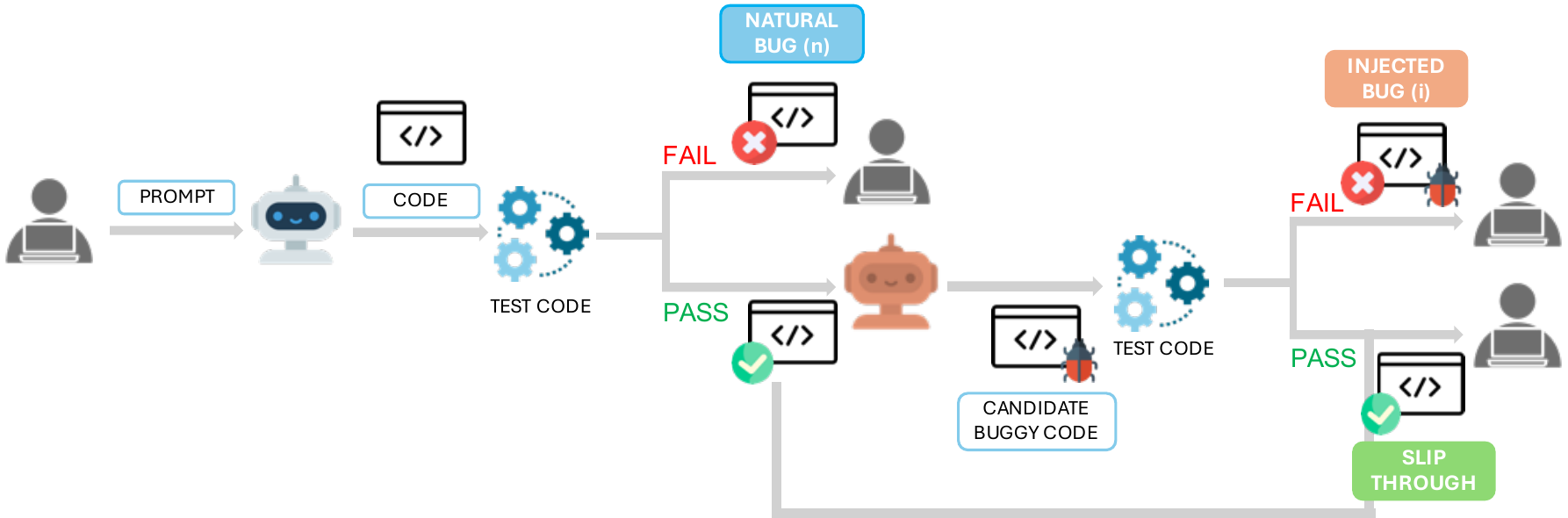}
     \caption{\looseness-1Bug injection pipeline in our study. After a student submits a prompt, the GenAI assistant generates code that is checked against hidden tests. Failing code is served unchanged as a natural bug (also referred to as `n'); passing code is routed through a bug injection step that proposes and validates candidate buggy code variants, and a validated failing variant is served as an injected bug (also referred to as `i'). If no candidate can be validated as a runnable failing variant, the original passing code is served unchanged, i.e., a slip-through.
     }
     \Description[Pipeline that turns passing code into a validated injected bug]{Left-to-right pipeline diagram with icons and labeled arrows. A student prompt goes to the GenAI assistant, which outputs code. The code is executed against hidden tests. If tests fail, the output is returned as a natural bug. If tests pass, the code is routed to a bug-injection step that produces candidate buggy code variants, which are tested again. A candidate that fails tests is served as an injected bug. If the candidate passes, the original solution is served as a slip-through. The diagram shows the decision points labeled PASS and FAIL and the three possible outcomes: natural bug, injected bug, and slip-through.}
    \label{fig.system}
\end{figure*}

Several tools have been reported in the literature for supporting prompt-based programming activities, sharing several similar features \cite{denny2023promptlyusingpromptproblems, padurean2025prompt, DBLP:conf/sigcse/YehTGYFC25}.  
To conduct our study, we used Prompt Programming \cite{padurean2025prompt}, our publicly available web platform that supports prompt-based programming through dialogue with a GenAI assistant, hidden-test execution, and direct code editing (see Figure~\ref{fig.illustration}). For this study, we extended the platform with a bug injection pipeline that can replace otherwise correct generated code with validated, runnable near-miss variants. In Prompt Programming, for each task students receive a problem specification and are required to craft a natural-language prompt to generate a C implementation that matches it. Once a response is received, students can execute the code against hidden unit tests to receive immediate feedback on correctness (including any failed test case) via console output. When a run fails, students can either continue the dialogue by sending another prompt or switch to the built-in editor to make code changes and re-run the edited code. Our GenAI assistant was backed by GPT-4o-mini, and the ``prompt-and-test'' workflow allows for an iterative approach to problem-solving where students alternate between specification and repair. Our platform also includes a pipeline for injecting bugs into code generated by this model, as we detail in the following section.
\subsection{Our Bug Injection Pipeline}\label{sec:method.pipeline}

\looseness-1To investigate student debugging behaviors, we incorporate our bug injection pipeline into the prompt-based programming workflow. This system creates deliberate, runnable failures, requiring students to engage in code verification and targeted repair, rather than relying only on prompt iteration to refine the specification. Students also encounter incorrect GenAI outputs during normal prompting. We refer to these naturally occurring prompt-related failures as natural bugs. These natural bugs often stem from specification mismatches or logic errors in the generated code. In contrast, authentic debugging scenarios also involve ``near-miss'' solutions, i.e., code that seems correct and close to a working solution, but contains a localized bug. We use a GenAI model for injection because it can propose context-aware edits that match the structure and style of the originally generated solution. We validate buggy candidates against the hidden tests to ensure the injected version remains executable and actually fails. Thus, our pipeline allows the existence of both natural and injected bugs during the same interaction, without disrupting the normal prompt-based programming workflow. 

\looseness-1The bug injection pipeline, illustrated in Figure~\ref{fig.system}, consists of multiple steps. It acts as a middleware layer between the GenAI assistant's code generation and the code shown to students. First, the student sends a prompt, and the GenAI generates candidate solution code. The pipeline intercepts the generated code and executes it against hidden tests. If the code fails, it is sent to the student unchanged and is considered to contain a \emph{natural bug (`n')}. If the code passes, it is sent back to the same GenAI model (GPT-4o-mini) under a different system prompt, where it acts as a ``bug injection expert'' and injects a realistic bug (examples shown in Table~\ref{tab.bug.examples}). The bug injection expert produces candidate buggy variants (we set the number of generated variants to $5$), which are then validated against the hidden test cases. We use the first variant that compiles, but fails the tests, to replace the original code shown to the student, which is then treated as containing an \emph{injected bug (`i')}. If all generated variants do not compile or pass the tests, bug injection is considered unsuccessful and the student is shown the original correct code. We refer to these cases as \emph{slip-throughs}. We limit bug injection to this set of candidates, skipping additional injection rounds and permitting some slip-through cases, since repeated attempts would slow responses and interfere with real-time interaction workflow. Importantly, because this middleware is applied to every assistant-generated code response, students generally cannot solve the task through prompting alone (except in slip-through cases) and must verify and repair code directly to complete the task.

\begin{figure*}[t]
\centering
    \begin{subfigure}{0.5\textwidth}
    \centering
        \includegraphics[width=\textwidth]{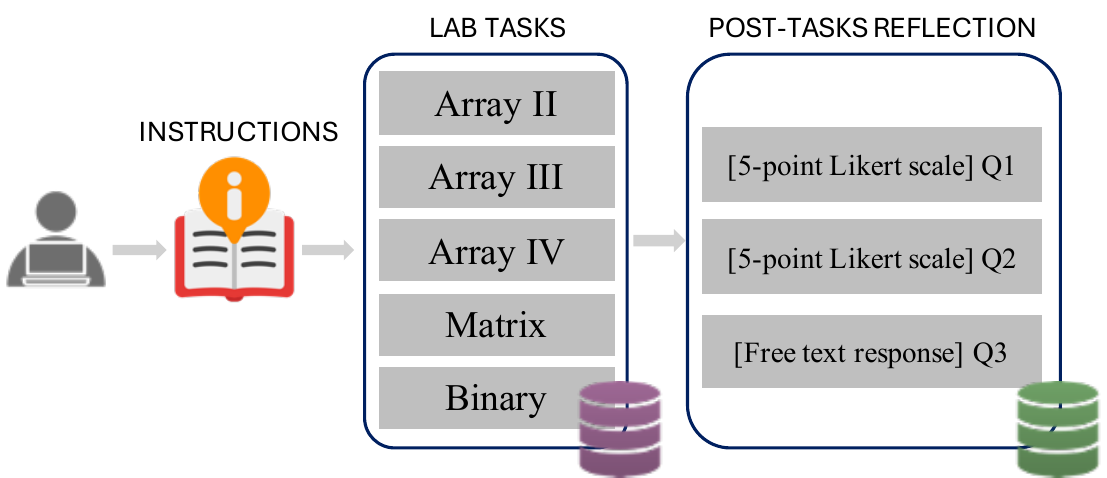}
        \caption{Data collection procedure}
        \label{fig.datacollection.flow}
    \end{subfigure}
    \begin{subfigure}{0.45\textwidth}
    \centering
        \scalebox{0.9}{
        \setlength\tabcolsep{4.0pt}
        \renewcommand{\arraystretch}{1.25}
        \begin{tabular}{@{} l >{\raggedright\arraybackslash}p{0.66\linewidth} @{}}
        \toprule
        \textbf{Task} & \textbf{Description} \\
        \midrule
        Array II   & Count negative values in an array \\
        Array III & Index of the last zero in an array\\
        Array IV      & Partially sorting an array within a specified index range \\
        Matrix          & Propagate $1$'s across the corresponding rows and columns in a 2D array (matrix) \\
        Binary & Compute fixed-size binary addition \\
        \bottomrule
        \end{tabular}
        }
    \caption{List of programming tasks}
    \label{fig.datacollection.tasks}
    \end{subfigure}
    \\
    \vspace{2mm}
    \begin{subfigure}{0.98\textwidth}
    \centering
    \scalebox{0.9}{
        \setlength\tabcolsep{4.0pt}
        \renewcommand{\arraystretch}{1.0}
        \begin{tabular}{@{} l >{\raggedright\arraybackslash}p{0.98\linewidth} @{}}
        \toprule
        \textbf{Question} & \textbf{Description} \\
        \midrule
        Q1   & [5-point Likert scale] I found it easy to locate and fix bugs in the code \\
        Q2 & [5-point Likert scale] I understand code I have written myself better than code generated by an AI model \\
        Q3  & [Free-text response] What benefits do you believe this version of the prompt-based programming task (i.e., where the model deliberately makes mistakes) offers for your learning? \\
        \bottomrule
        \end{tabular}
        }
        \caption{Post-tasks reflection questionnaire}
        \vspace{-2mm}
    \label{fig.datacollection.reflection}
    \end{subfigure}
    \caption{\looseness-1Classroom deployment. (a) presents the two stage data collection procedure in the study. In the first stage, data is collected from platform logs. In the second stage, we collect student reflections after completing the tasks. (b) details the five programming tasks used for the study. (c) presents details of the post-tasks reflection questions. Details of data collection presented in Section~\ref{sec:method.collection}.}
    \Description[Classroom deployment overview with tasks and reflection questions]{Three-part figure. Panel (a) shows an activity flow: students receive instructions, complete several lab tasks on the platform, and generate logged data, followed by a post-task reflection step. Panel (b) is a table listing five tasks, Array II, Array III, Array IV, Matrix, and Binary, each with a one-sentence description of the programming goal. Panel (c) is a table listing three reflection questions: two 5-point Likert items about bug-fixing ease and understanding self-written versus AI-generated code, plus one free-text question about perceived learning benefits.}
    \label{fig.datacollection}
\end{figure*}

\begin{figure*}[t!]
\centering
	\includegraphics[width=\textwidth]{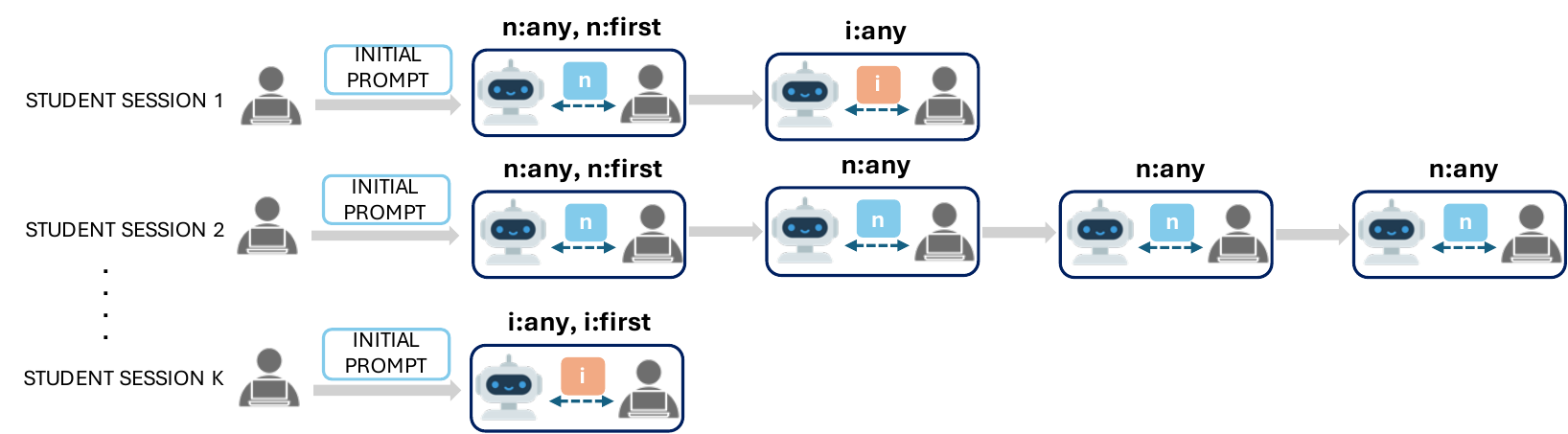
    }
     \caption{\looseness-1Examples of student sessions, along with the terminology used throughout analysis. A session consists of a sequence of student–GenAI interactions, and each turn is defined as a student response to a preceding GenAI message. The GenAI response in a turn may contain a natural bug (n) or an injected bug (i). We label turns containing natural bugs as \textbf{n:any} and turns containing injected bugs as \textbf{i:any}. We additionally label the first turn in a session, when it contains buggy code, as \textbf{n:first} or \textbf{i:first}, respectively. Session 1 corresponds to the scenario in Figure~\ref{fig.illustration} and contains two turns: the first turn is labeled \textbf{n:any} and \textbf{n:first}, and the second turn is labeled \textbf{i:any}. Session 2 shows multiple consecutive natural-bug turns, with only the first labeled \textbf{n:first}. Session K consists of a single injected bug turn, labeled both \textbf{i:any} and \textbf{i:first}. All behavioral analyses are aggregated at the turn level. Additional details are in Section~\ref{sec:method}.}
     \Description[Examples of sessions and labels n:any, n:first, i:any, i:first]{Schematic timelines of student sessions showing alternating student and assistant icons. Each session begins with an initial prompt and then turns where first an assistant message comes up and the student is responding to that message. Bug source is indicated with letters n for natural and i for injected, and labels appear above the assistant responses. Session 1 contains two buggy turns: the first labeled n:any and n:first, followed by an injected turn labeled i:any. Session 2 shows multiple consecutive natural-bug turns labeled n:any, with only the first also labeled n:first. Session K shows a single injected-bug turn labeled i:any and i:first.}
    \label{fig.session}
\end{figure*}

\subsection{Classroom Deployment}\label{sec:method.collection}

\looseness-1Data was collected from an introductory C programming course (CS1) at the University of Auckland during the second semester of 2025, under the approval of the university human ethics committee. The overall data collection workflow is shown in Figure~\ref{fig.datacollection.flow}. The study was conducted as part of a graded take-home lab assignment which could be completed within a one-week window. Students worked on five prompt-based programming tasks, namely Array II, Array III, Array IV, Matrix, and Binary (task details are presented in Figure~\ref{fig.datacollection.tasks}). Tasks covered core CS1 topics such as arrays, matrices, and binary operations. At the start of the activity, students were explicitly warned that the code generation model was likely to make mistakes, then instructed to carefully read generated code and manually edit it where needed, framing the lab as verification and repair practice alongside prompting. During the activity, the interface did not indicate whether failing code was due to a natural or injected bug. Students received credit for solving at least two of the five problems. Data was collected in two stages. First, the platform recorded interaction logs during student attempts, including prompts, generated code, edits, and code executions. Second, immediately after completing the tasks, students completed a reflection questionnaire designed to capture their experiences with code understanding and debugging (reflection questions are shown in Figure~\ref{fig.datacollection.reflection}). Of $1037$ enrolled students, $917$ engaged with at least one task and are included in this analysis; $897$ of these $917$ students ($97.82\%$) solved at least two problems and thus met the credit threshold.

\begin{figure*}[t]
\centering
    \includegraphics[width=0.99\textwidth]{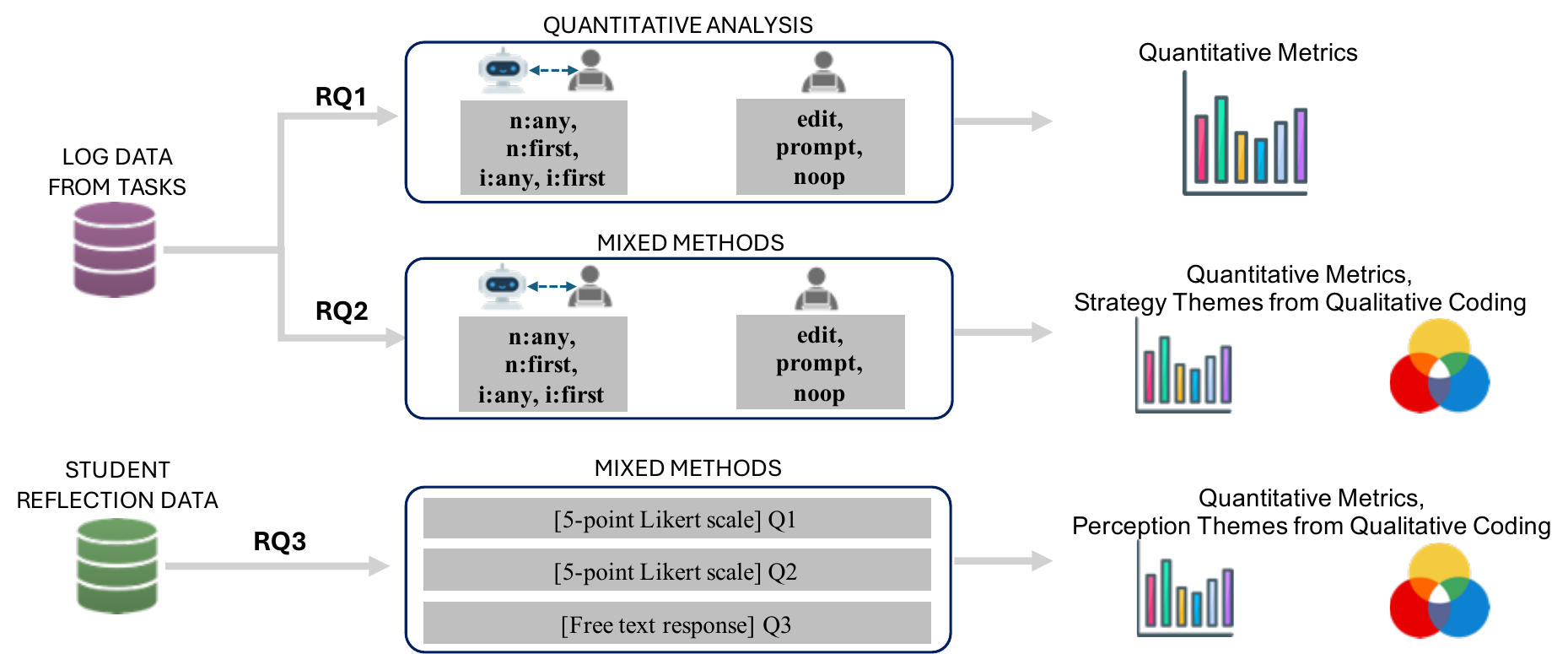}
    \vspace{3mm}
    \caption{\looseness-1Data analysis procedures. Overview of the data analysis procedures across research questions. (top) For RQ1 and RQ2, we analyze log data collected from students’ interactions with the programming tasks. Analyses are conducted at the turn level (n:any, n:first, i:any, i:first) and examine subsequent student actions (any, noop, edit, prompt). We perform quantitative analyses to measure behavioral patterns and outcomes, and qualitative coding to identify student strategies for responding to natural and injected bugs. (bottom) For RQ3, we analyze student reflection data collected at the end of the lab, including Likert-scale responses (e.g., perceived ease of bug fixing and code comprehension) and free-text responses. Quantitative analyses summarize Likert responses, while thematic coding extracts themes on perceived learning benefits from free-text reflections. Further details on data analysis procedure are presented in Section~\ref{sec:method.analysis}.}
    \Description[Analysis workflow mapping platform logs and reflections to RQ1–RQ3]{Block diagram linking data sources to analyses and outputs. Log data from tasks flows into RQ1 and RQ2 analyses; reflection data flows into RQ3. RQ1 is shown as quantitative analysis at the turn level using labels n:any, n:first, i:any, and i:first and first actions edit, prompt, or noop, producing quantitative metrics. RQ2 is shown as mixed methods, combining the same quantitative metrics with qualitative coding to derive strategy themes. RQ3 is shown as mixed methods over Likert and free-text responses, producing quantitative summaries and qualitative perception themes.}
    \label{fig.datanalysis}
\end{figure*}

\subsection{A Student Session in the Platform}\label{sec:method.session}
\looseness-1Our analysis is based on the platform's event logs, which record student–GenAI message exchanges, code edits, execution history, and a bug injection audit trail. The audit log allows us to verify whether each GenAI solution was successfully modified with an injected bug or left unchanged. To reduce noise typical of web-based systems (e.g., timeouts or interrupted sessions), we apply conservative filtering and exclude sessions that do not follow strict alternation between student and assistant messages (i.e., prompts and GenAI replies must alternate in timestamp order, with each prompt followed by a reply; we did not apply any time-gap threshold) or for which no executable code artifact can be recovered. This removed $177$ ($6.3\%$) of $2{,}813$ logged sessions, leaving $2{,}636$ for our analysis.

\looseness-1In our analysis, we define a \emph{`session'} as one student's sequential interaction with the GenAI assistant while attempting to solve one problem, ending when the student either succeeds or disengages from the task. There is at most one session for a student-problem pair. A \emph{`turn'} is defined as a student's follow-up action in response to a specific GenAI message, conditioned on the state of the immediately preceding GenAI message. Specifically, we categorize the preceding GenAI message as: (1) no-code (the message contains no code); (2) natural bug `n' (the generated code contains a natural bug); (3) injected bug `i' (the generated code contains an injected bug); and (4) slip-through (generated code is correct and passes all test cases). No-code messages typically occur when the assistant responds with clarification questions, explanations, or other non-code guidance rather than an identifiable program. For turn-level analyses, we use two complementary labels. The labels \textbf{n:any} and \textbf{i:any} denote any turn in which the student is responding to a natural or injected bug, respectively. We also mark the first turn in each session, when it contains buggy code, as \textbf{n:first} or \textbf{i:first}, which lets us analyze students' initial response to failure before later turns in the same session introduce repeated exposure or strategy carryover. Figure~\ref{fig.session} illustrates three example sessions and these labels. For behavior and outcome analyses, we focus on turns where the student is responding to buggy code (natural or injected bug). Within each turn, we operationalize the student's next-step response using the first logged follow-up action, i.e., \emph{prompt} if the student submits a new prompt before making any manual edits, and \emph{edit} if the student edits code before sending a follow-up prompt. Turns where neither occurs are labeled as \emph{noop}.

\subsection{Data Analysis Procedures}\label{sec:method.analysis}

Our analysis procedure is summarized in Figure~\ref{fig.datanalysis}, which shows how different data sources support answering RQs. In particular, RQ1 and RQ2 analyze student log data from the platform, while RQ3 analyzes the post-tasks reflection data.

\looseness-1For RQ1 and RQ2, we analyze student behavior at the turn level, focusing on responses to natural and injected bugs. Within each buggy turn (i.e., excluding slip-throughs), we classify actions (prompt, edit, or noop) and measure immediate success conditional on that action. For turns where students choose to prompt, immediate success is defined counterfactually using the injection audit, and a turn is marked successful if the subsequent assistant output \emph{would} have passed tests prior to bug injection. For turns where students edit, immediate success reflects whether the student obtains working code through only edit-and-run attempts without submitting another prompt. 
Because injected bugs are introduced only after an initially correct solution, comparisons between natural and injected bugs are interpreted as differences in response behavior rather than causal effects of bug source. For statistical testing, we restrict attention to the first buggy instance per session (\textbf{n:first} and \textbf{i:first}) to better satisfy independence assumptions. Each session therefore contributes at most one observation per problem. Testing results are reported at the per-problem level, and pooled summaries are descriptive. As tests are exploratory, we report uncorrected $p$-values. We use Pearson's $\chi^2$ test for categorical outcomes and Mann–Whitney tests for continuous measures.

\begin{table*}[t]
\caption{Descriptive statistics by problem. The table reports participation, session-level outcomes, and turn-level coverage conditioned on the preceding GenAI message, for all turns (any turn) and first turns only (first turn). Turn-level categories are no-code (the GenAI message did not contain code), natural bug, injected bug, and slip-through. The total turns ($6{,}706$) include no-code and slip-through turns, whereas the bug-fixing turns ($6{,}071$) include only natural and injected bug turns.}
\label{tab.stats}
\centering

\scalebox{0.97}{
\setlength\tabcolsep{3pt}
\renewcommand{\arraystretch}{1.15}
\begin{tabular}{l r r r r r r r r r r r r r}
\toprule
\multirow{3}{*}{\textbf{Problem}} &
\multirow{3}{*}{\begin{tabular}{c}\textbf{Nb.}\\\textbf{students}\end{tabular}} &
\multicolumn{2}{c}{\textbf{Session-level}} &
\multicolumn{5}{c}{\textbf{Turn-level: any turn}} &
\multicolumn{5}{c}{\textbf{Turn-level: first turn}} \\
\cmidrule(lr){3-4}\cmidrule(lr){5-9}\cmidrule(lr){10-14}
& &
\multicolumn{1}{c}{\textbf{Nb.}} &
\multicolumn{1}{c}{\textbf{Perc.}} &
\multicolumn{1}{c}{\textbf{Nb.}} &
\multicolumn{1}{c}{\textbf{No}} &
\multicolumn{1}{c}{\textbf{Natural}} &
\multicolumn{1}{c}{\textbf{Injected}} &
\multicolumn{1}{c}{\textbf{Slip}} &
\multicolumn{1}{c}{\textbf{Nb.}} &
\multicolumn{1}{c}{\textbf{No}} &
\multicolumn{1}{c}{\textbf{Natural}} &
\multicolumn{1}{c}{\textbf{Injected}} &
\multicolumn{1}{c}{\textbf{Slip}} \\
& &
\multicolumn{1}{c}{\textbf{sessions}} &
\multicolumn{1}{c}{\textbf{succ.}} &
\multicolumn{1}{c}{\textbf{turns}} &
\multicolumn{1}{c}{\textbf{code}} &
\multicolumn{1}{c}{\textbf{bug}} &
\multicolumn{1}{c}{\textbf{bug}} &
\multicolumn{1}{c}{\textbf{through}} &
\multicolumn{1}{c}{\textbf{turns}} &
\multicolumn{1}{c}{\textbf{code}} &
\multicolumn{1}{c}{\textbf{bug}} &
\multicolumn{1}{c}{\textbf{bug}} &
\multicolumn{1}{c}{\textbf{through}} \\
\midrule
Array II &
$$900$$ & $$900$$ & $$99.00\%$$ &
$$1848$$ & $$3.03\%$$ & $$31.17\%$$ & $$50.70\%$$ & $$15.10\%$$ &
$$900$$ & $$2.44\%$$ & $$33.33\%$$ & $$49.78\%$$ & $$14.44\%$$ \\
Array III &
$$866$$ & $$866$$ & $$97.00\%$$ &
$$2348$$ & $$1.32\%$$ & $$37.61\%$$ & $$59.16\%$$ & $$1.92\%$$ &
$$866$$ & $$1.39\%$$ & $$42.73\%$$ & $$55.08\%$$ & $$0.81\%$$ \\
Array IV &
$$386$$ & $$386$$ & $$76.94\%$$ &
$$1040$$ & $$2.12\%$$ & $$58.46\%$$ & $$29.42\%$$ & $$10.00\%$$ &
$$386$$ & $$1.81\%$$ & $$68.13\%$$ & $$21.76\%$$ & $$8.29\%$$ \\
Matrix &
$$270$$ & $$270$$ & $$82.96\%$$ &
$$607$$ & $$1.32\%$$ & $$42.17\%$$ & $$47.12\%$$ & $$9.39\%$$ &
$$270$$ & $$1.11\%$$ & $$45.56\%$$ & $$45.93\%$$ & $$7.41\%$$ \\
Binary &
$$214$$ & $$214$$ & $$50.93\%$$ &
$$863$$ & $$3.36\%$$ & $$89.34\%$$ & $$6.84\%$$ & $$0.46\%$$ &
$$214$$ & $$3.27\%$$ & $$92.99\%$$ & $$3.74\%$$ & $$0.00\%$$ \\
\midrule
All Probl. &
$$917$$ & $$2636$$ & $$89.57\%$$ &
$$6706$$ & $$2.18\%$$ & $$46.14\%$$ & $$44.39\%$$ & $$7.29\%$$ &
$$2636$$ & $$1.93\%$$ & $$47.61\%$$ & $$43.29\%$$ & $$7.17\%$$ \\
\bottomrule
\end{tabular}
}

\end{table*}

\looseness-1To analyze students' prompting strategies during bug-fixing turns, we use a single-label coding scheme that assigns each sampled prompt one primary strategy (i.e., student's approach to design a follow-up prompt to resolve the existing bug), following standard content analysis practice \cite{Neuendorf2002ContentAnalysisGuidebook,Stemler2000OverviewContentAnalysis}. We sampled prompts uniformly from bug-fixing turns where students chose to prompt, making sure that each student is represented only once. Two annotators followed a two-phase process. In the open coding phase, they independently coded $60$ prompts to identify strategy categories and construct an inductive codebook. In the closed coding phase, they first calibrated on the open-coded examples, then independently coded $144$ additional prompts. We estimated inter-rater reliability (IRR) using Krippendorff's $\alpha$, treating values above $0.8$ as reliable \cite{Krippendorff2011ComputingKA,krippendorff2018content}. After computing IRR ($\alpha=0.87$), the annotators reconciled disagreements, and we use the reconciled labels for reporting. No new strategy categories were introduced during closed coding and reconciliation, suggesting the codebook had stabilized for the sampled prompts. In total, $204$ prompts were coded.

\begin{table*}[t]
\caption{Examples of injected bugs by problem. We also report the number of distinct injected bugs for each problem after normalizing away comments, formatting, and variable namings, while keeping logic-changing differences such as operators and numbers. We show two representative examples chosen from the five most frequent injected bugs for that problem. Each example is presented as a compact changed-line diff, where the first line shows the original code (highlighted in {\setlength{\fboxsep}{1pt}\colorbox{diffold}{light green}}
) and the second line shows the buggy version (highlighted in {\setlength{\fboxsep}{1pt}\colorbox{diffnew}{light red}}), followed by a short explanation of what was changed.}
\label{tab.bug.examples}
\centering

\scalebox{0.97}{
\setlength\tabcolsep{10pt}
\renewcommand{\arraystretch}{1.05}
\begin{tabular}{c c c}
\toprule
\textbf{Problem} & \textbf{Nb. bugs} &
\begin{tabular}[c]{@{}p{5.8cm}p{6.8cm}@{}}
\textbf{\ \ Example bug diff} & \textbf{Explanation}
\end{tabular} \\
\midrule

Array II &
$104$ &
\begin{tabular}[c]{@{}p{5.8cm}p{6.8cm}@{}}
\begin{tabular}[t]{@{}l@{}}
\ \ \oldline{int count = 0;} \\
\ \ \bugline{int count;}
\end{tabular}
&
The counter variable is no longer initialized to $0$, so it starts with an indeterminate value. \\
\cmidrule(lr){1-2}
\begin{tabular}[t]{@{}l@{}}
\ \ \oldline{count++;} \\
\ \ \bugline{count += 2;}
\end{tabular}
&
Each negative value is counted twice.
\end{tabular}
\\
\midrule

Array III &
$112$ &
\begin{tabular}[c]{@{}p{5.8cm}p{6.8cm}@{}}
\begin{tabular}[t]{@{}l@{}}
\ \ \oldline{for (int i = size - 1; i >= 0; i-{}-)} \\
\ \ \bugline{for (int i = size; i >= 0; i-{}-)}
\end{tabular}
&
The loop starts one position past the valid range. \\
\cmidrule(lr){1-2}
\begin{tabular}[t]{@{}l@{}}
\ \ \oldline{return i;} \\
\ \ \bugline{return i + 1;}
\end{tabular}
&
The index returned after finding the last zero is shifted by one.
\end{tabular}
\\
\midrule

Array IV &
$112$ &
\begin{tabular}[c]{@{}p{5.8cm}p{6.8cm}@{}}
\begin{tabular}[t]{@{}l@{}}
\ \ \oldline{for (int j = i + 1; j <= end; j++)} \\
\ \ \bugline{for (int j = start; j <= end; j++)}
\end{tabular}
&
The inner loop of the sorting logic wrongly starts from the beginning of the target subarray (also in Figure~\ref{fig.illustration}). \\
\cmidrule(lr){1-2}
\begin{tabular}[t]{@{}l@{}}
\ \ \oldline{if (array[i] > array[j])} \\
\ \ \bugline{if (array[i] < array[j])}
\end{tabular}
&
The comparison direction for sorting is flipped, pushing elements in the wrong order.
\end{tabular}
\\
\midrule

Matrix &
$\phantom{0}69$ &
\begin{tabular}[c]{@{}p{5.8cm}p{6.8cm}@{}}
\begin{tabular}[t]{@{}l@{}}
\ \ \oldline{for (int i = 0; i < rows; i++)} \\
\ \ \bugline{for (int i = 0; i <= rows; i++)}
\end{tabular}
&
The loop runs one step past the valid range when iterating over matrix rows. \\
\cmidrule(lr){1-2}
\begin{tabular}[t]{@{}l@{}}
\ \ \oldline{rowFlags[i] = 1; colFlags[j] = 1;} \\
\ \ \bugline{rowFlags[j] = 1; colFlags[i] = 1;}
\end{tabular}
&
The row and column indices are swapped when marking flags for propagation.
\end{tabular}
\\
\midrule

Binary &
$\phantom{0}43$ &
\begin{tabular}[c]{@{}p{5.8cm}p{6.8cm}@{}}
\begin{tabular}[t]{@{}l@{}}
\ \ \oldline{int carry = 0;} \\
\ \ \bugline{int carry = 1;}
\end{tabular}
&
The carry variable is initialized to $1$ instead of $0$, introducing a spurious initial carry. \\
\cmidrule(lr){1-2}
\begin{tabular}[t]{@{}l@{}}
\ \ \oldline{carry = sum / 2;} \\
\ \ \bugline{carry = sum \% 2;}
\end{tabular}
&
Carry propagation for the binary addition uses the remainder instead of overflow.
\end{tabular}
\\
\bottomrule
\end{tabular}
}

\end{table*}

\looseness-1For RQ3, we analyze post-tasks reflections using both quantitative summaries of Likert-scale responses and multi-label thematic analysis of responses to the free-text question \cite{BraunClarke2006Thematic,ClarkeBraun2017Thematic}. For thematic analysis, we sampled reflections uniformly, as each student contributes at most one reflection about the activity. Thematic coding followed open and closed phases, with annotators coding $54$ responses in open coding and an additional $150$ responses in closed coding. Reliability was assessed using Krippendorff's $\alpha$ per label, computed by binarizing label presence/absence per response (mean $\alpha = 0.82$, range $=0.59$-$0.96$) \cite{Krippendorff2011ComputingKA,krippendorff2018content,csanyi2025sentence}, and MASI-based agreement ($\alpha = 0.80$) \cite{DBLP:conf/lrec/Passonneau06,DBLP:conf/icer/DemirtasFHC24}. No new themes emerged during closed coding and reconciliation. We report results using $204$ reconciled responses.



\section{Results}
\label{sec:results}

In this section, we first show descriptive statistics of the collected data, and then present results of our study centered around the three research questions formulated in Section~\ref{sec:introduction}.

\subsection{Descriptive Statistics}

\begin{figure*}[t!]
  \centering
  \captionsetup[subfigure]{justification=centering,singlelinecheck=false}
  
    \begin{subfigure}[t]{0.99\linewidth}
      \centering
      
      \includegraphics[width=0.8\linewidth,trim={2mm 4mm 2mm 5mm},clip]{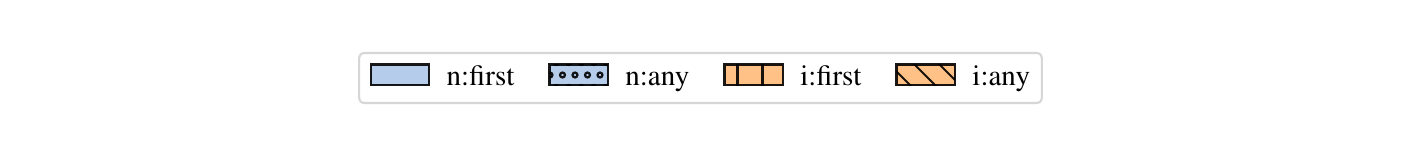}
      \vspace{-2.5mm}
      
      \includegraphics[width=0.98\linewidth,trim={0mm 0mm 0mm 2mm},clip]{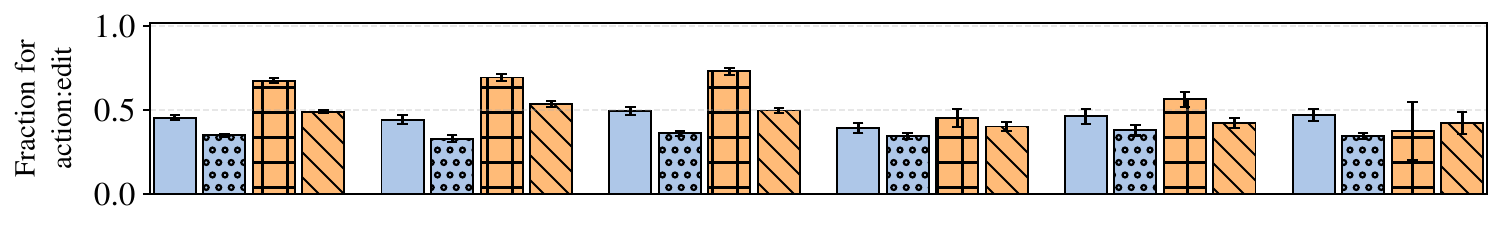}
      \vspace{-1mm}
      
      \includegraphics[width=0.985\linewidth,trim={0mm 0mm 0mm 2mm},clip]{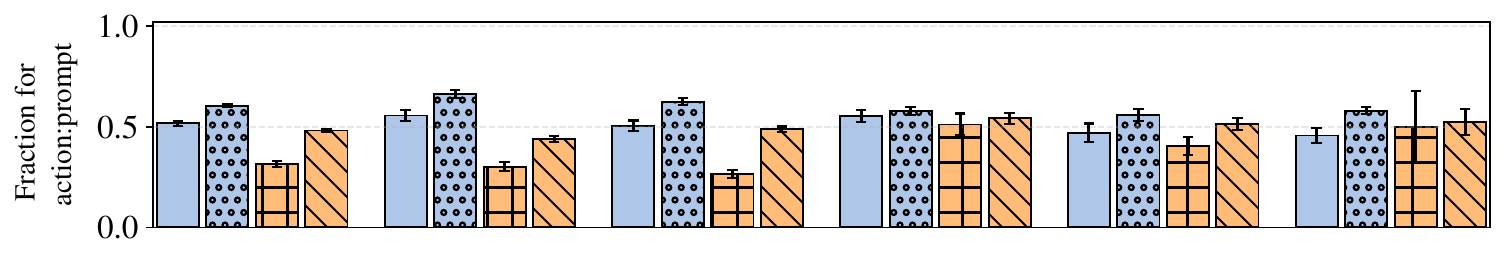}
      \vspace{-1mm}
      
      \includegraphics[width=0.992\linewidth,trim={0mm 0mm 0mm 2mm},clip]{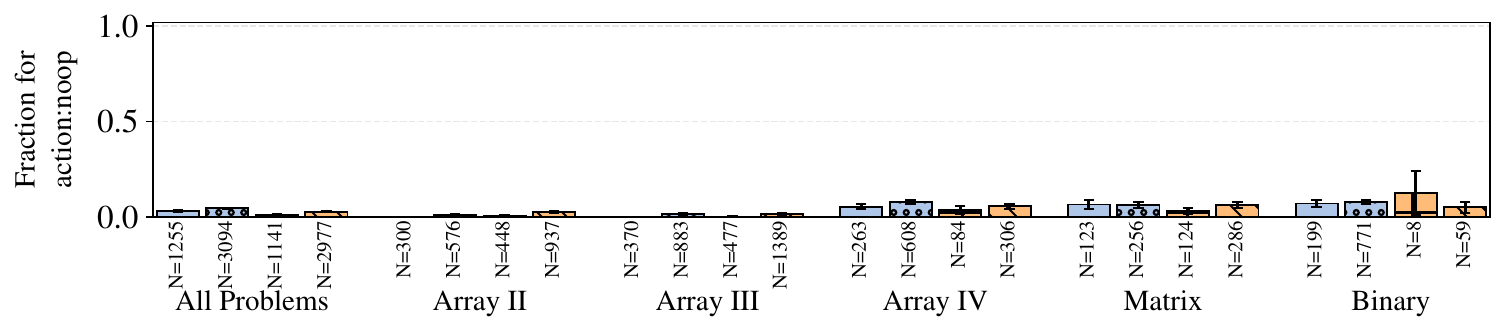}
      \caption{Fraction of turns where students chose to edit first (top), prompted (middle), or did not follow up with an action (bottom).}
      \label{fig.rq1.category.actions}
    \end{subfigure}
  \\
  \vspace{3.5mm}
  \begin{subfigure}[t]{0.9\linewidth}
    \centering
    \includegraphics[width=0.99\linewidth]{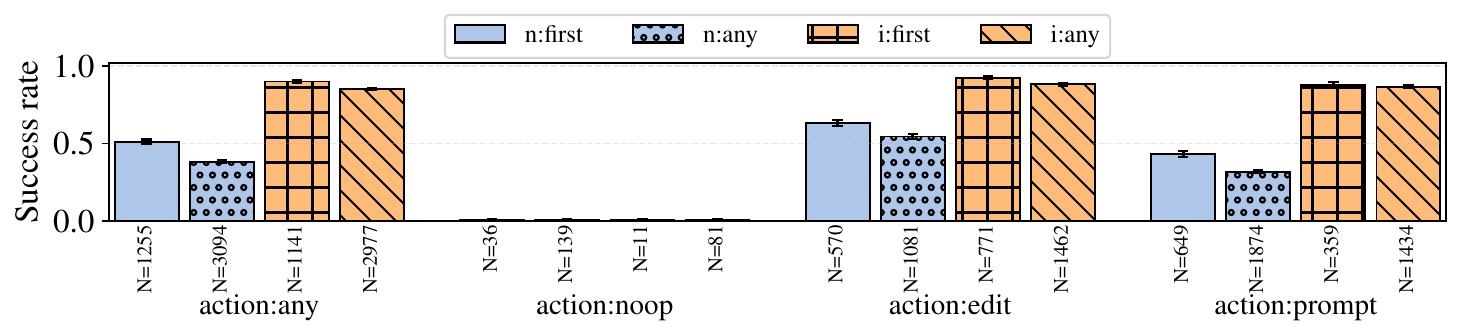}
    \caption{Fraction of turns with immediate success, conditioned on students' actions.}
    \label{fig.rq1.category.success}
  \end{subfigure}
    \caption{RQ1 quantitative results. (a) shows the first-action distributions after a buggy GenAI response (shown for four source-based groups \textbf{n:first}, \textbf{n:any}, \textbf{i:first}, and \textbf{i:any}); fractions across the three action categories sum to $1$ within source-based group. (b) shows the immediate success rates for the same groupings, conditioned by action. For cases where students prompted, immediate success is defined using the pre-injection audit outcome of the next generated code; for cases where the student edited, it is defined by reaching a passing result through edit-and-run attempts before the next prompt; `noop' turns naturally have $0$ immediate success.}
    \Description[RQ1 results comparing first actions and immediate success across bug sources]{Two-panel figure of grouped bar charts. Panel (a) contains three stacked plots with the same x-axis categories: All Problems, Array II, Array III, Array IV, Matrix, and Binary. The y-axes are fractions for first action: edit, prompt, and noop. Each category shows four bars corresponding to n:first, n:any, i:first, and i:any. The edit-first fraction is higher for injected-bug groups than natural-bug groups, while the prompt-first fraction is higher for natural-bug groups. Noop fractions are small across categories. Panel (b) shows success rates grouped by action (any, noop, edit, prompt) for the same four bug-source groups, with injected-bug groups showing higher immediate success than natural-bug groups for prompt and edit actions.}
    \label{fig.rq1}
\end{figure*}

\looseness-1Table~\ref{tab.stats} presents an overview of participation, session-level, all-turns-level, and first-turn-level statistics. The ``Natural bug'' and ``Injected bug'' columns under ``Turn-level: any turn'' correspond to \textbf{n:any} and \textbf{i:any}, respectively, while the ``Natural bug'' and ``Injected bug'' columns under ``Turn-level: first turn'' correspond to \textbf{n:first} and \textbf{i:first}. Across all problems, the dataset includes $917$ participating students and $2{,}636$ sessions, with an overall session success rate of $89.57\%$. Session success and participation were highest on Array II and Array III and lowest on Binary, with Matrix and Array IV in between. Across all problems, natural and injected bug sources were broadly balanced in both the `any turn' and `first turn' levels. However, their distribution varied widely by problem.

\looseness-1Table~\ref{tab.bug.examples} shows examples of injected bugs observed in practice, collected from our bug audit logs. For each problem, we report the number of distinct bugs after grouping together superficial variants of the same change. We do this by comparing only the changed lines between the original and injected code, ignoring comments and formatting, and merging simple identifier renamings, while keeping logic-changing edits such as operator and numeric differences separate. Overall, the examples suggest that injected bugs are typically small, local mutations, the majority of which resemble plausible near-miss faults that students may realistically encounter in practice when reading, testing, and repairing code.

\subsection{RQ1: Follow-Up Action Choices and Immediate Outcomes After Bugs}

To address our first RQ, we analyze students' actions and immediate success rates after buggy GenAI responses, conditioned on bug source. Figure~\ref{fig.rq1} presents an overview of the results.

\looseness-1\textbf{Follow-up action choice.} Specifically, Figure~\ref{fig.rq1.category.actions} reports student actions after a buggy GenAI response, conditioned on bug source (i.e., \textbf{n:first}, \textbf{i:first}, \textbf{n:any}, and \textbf{i:any}). We show results both pooled across problems and broken down per problem. Pooled across all problems, the `first turn' summaries show a clear shift by bug source. Students are more likely to choose prompting in \textbf{n:first} than in \textbf{i:first} ($51.71\%$ vs.\ $31.46\%$), while edit actions are more common in \textbf{i:first} than in \textbf{n:first} ($67.57\%$ vs.\ $45.42\%$). The same trend appears in the `any turn' summaries, with \textbf{n:any} having a higher share of prompt actions than \textbf{i:any} ($60.57\%$ vs.\ $48.17\%$), while \textbf{i:any} has a higher share of edit actions than \textbf{n:any} ($49.11\%$ vs.\ $34.94\%$). Noop responses, where students neither reprompt nor edit, are rare. The per-problem breakdown follows the same overall trend, with the clear contrasts in Array II and Array III, but smaller gaps in Array IV and Matrix. Binary has sparser coverage, making comparison noisier\footnote{Per problem, we compare \textbf{n:first} vs.\ \textbf{i:first} action distributions using $\chi^2$. The directional difference was strongly supported in Array II ($\chi^2(2)=49.40$, $p<.001$) and Array III ($\chi^2(2)=51.57$, $p<.001$), but not in Array IV ($\chi^2(2)=1.20$, $p=.550$) or Matrix ($\chi^2(2)=3.25$, $p=.197$). Binary too sparse.}.

The results suggest that injected bugs are often encountered as locally repairable near-misses. Relative to natural bugs, they are associated with less prompting and more editing. This pattern is consistent with injected bugs being introduced only after the assistant first produces correct code, necessitating from students a smaller and more localized repair and creating more direct opportunities for verification and repair.

\textbf{Immediate success rates.} Figure~\ref{fig.rq1.category.success} reports immediate success rates after a buggy GenAI response, conditioned on the student's first action and bug source. We focus on the overall trends and therefore show pooled rates across problems. Immediate success rates following student responses to injected bugs are higher than those after natural bugs. In the `first turn' summaries, success for any action is $89.92\%$ for \textbf{i:first} vs. $51.16\%$ for \textbf{n:first}; the same contrast holds within prompt actions ($87.74\%$ vs.\ $43.30\%$) and edit actions ($92.22\%$ vs.\ $63.33\%$). In the `any turn' summaries, success is $84.95\%$ for \textbf{i:any} vs. $38.30\%$ for \textbf{n:any}, with the same trends for prompt ($86.54\%$ vs.\ $31.80\%$) and edit actions ($88.10\%$ vs.\ $54.49\%$). Under the \textbf{n:first} vs.\ \textbf{i:first} conditions, immediate success differs by bug source for any action in Array II, Array III, Array IV, and Matrix ($\chi^2$, df $=1$, all $p<.001$); Binary interpreted descriptively due to sparse data. The same trends hold within prompt and edit actions, with $p<.001$ in most problems (except for prompt action in Array IV $p=.003$, and edit action in Array II $p=.038$).\footnote{Corresponding $\chi^2(1)$ statistics (Array II, Array III, Array IV, Matrix) are: any action ($66.50$, $88.96$, $34.24$, $44.43$); prompt ($46.39$, $45.35$, $9.08$, $30.23$); edit ($4.32$, $31.05$, $25.55$, $11.54$). Binary is too sparse for reliable testing.}

The immediate success pattern aligns with intervention design. In particular, injected bugs are introduced only after the assistant first produces a correct solution, leaving students to repair a localized fault. Natural bugs, in contrast, more often reflect specification or logic mismatches, making immediate resolution less likely even after a reasonable action.

\looseness-1\textbf{RQ1 summary.} Overall, our results highlight differences by bug source in both follow-up actions and immediate success. We observe that injected bugs more often come with editing rather than reprompting, and are more likely to be resolved immediately. We do not interpret this as a causal difficulty comparison, because injected bugs are introduced only after the GenAI first produces a correct solution. Instead, these patterns suggest that bug source is associated with different response contexts and likely next actions. This comes in line with the intended design, i.e., natural bugs more often coincide with specification refinement, while injected bugs more often coincide with verification and localized repair.

\subsection{RQ2: How Prompting and Editing Strategies Differ According to Bug Sources}

\begin{figure*}[t!]
  \centering
  \captionsetup[subfigure]{justification=centering,singlelinecheck=false}

  \includegraphics[width=0.8\linewidth,trim={2mm 0mm 2mm 0mm},clip]{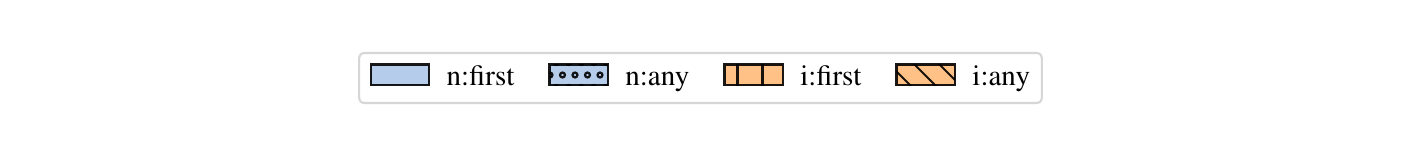}
  \vspace{-4mm}

  \begin{subfigure}[t]{0.32\linewidth}
    \centering
    \includegraphics[width=\linewidth]{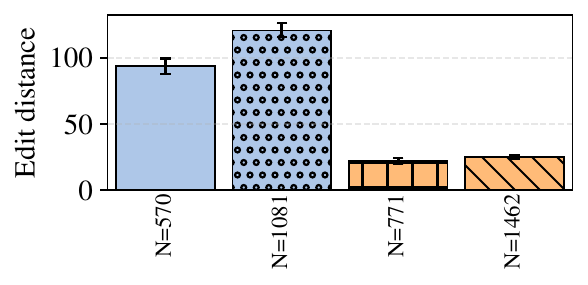}
    \vspace{-6mm}
    \caption{Edit distance}
    \label{fig.rq2.a}
  \end{subfigure}\hfill
  \begin{subfigure}[t]{0.305\linewidth}
    \centering
    \includegraphics[width=\linewidth]{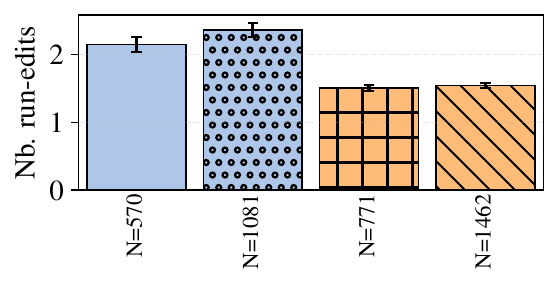}
    \vspace{-6mm}
    \caption{Edit-and-run count}
    \label{fig.rq2.b}
  \end{subfigure}\hfill
  \begin{subfigure}[t]{0.32\linewidth}
    \centering
    \includegraphics[width=\linewidth]{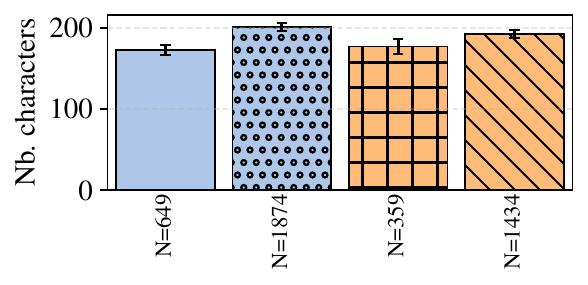}
    \vspace{-6mm}
    \caption{Prompt length}
    \label{fig.rq2.c}
  \end{subfigure}

  \par\vspace{5mm}

  \begin{subfigure}[t]{0.55\linewidth}
    \vspace{0pt}
    \centering
        \scalebox{0.93}{
        \setlength\tabcolsep{4.0pt}
        \renewcommand{\arraystretch}{1.53}
        \begin{tabular}{@{} l >{\raggedright\arraybackslash}p{0.66\linewidth} @{}}
        \toprule
        \textbf{Strategy} & \textbf{Primary intent} \\
        \midrule
        Task reframe   & Reframe the specification, reinterpret the goal of the task. \\
        Update signature & Change the function signature or arguments (add, remove, rename). \\
        Extend task      & Add constraints, details, or edge cases to refine the task. \\
        Restate          & Repeat the problem statement with minor tweaks, without new constraints. \\
        Provide guidance & Request a direct fix, or give imperative guidance on the approach. \\
        Provide output   & Provide an example, test case, or observed error / console output. \\
        Provide code     & Provide a full solution attempt (complete code). \\
        Meta             & Comment on the interaction (e.g., frustration) rather than the task. \\
        \bottomrule
        \end{tabular}
        }
    \caption{Prompting strategy codebook.}
    \label{fig.rq2.d}
  \end{subfigure}\hfill
  \begin{subfigure}[t]{0.43\linewidth}
    \vspace{0pt}
    \centering
    \includegraphics[width=\linewidth]{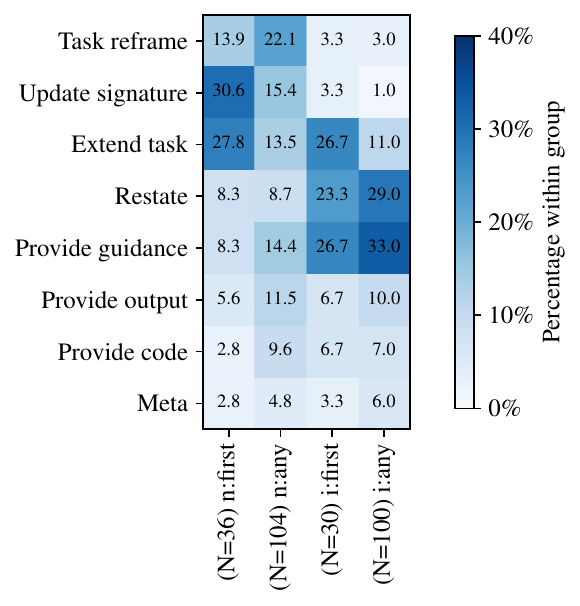}
    \vspace{-6mm}
    \caption{Prompting strategy distribution by bug source.}
    \label{fig.rq2.e}
  \end{subfigure}

  \caption{\looseness-1RQ2 mixed-method results. (a) shows the mean absolute Levenshtein distance between the initial buggy GenAI code and the student's last edited code, computed for cases where students chose an edit action. (b) shows the mean number of edit-and-run attempts within the same cases. (c) shows the mean prompt length in number of characters for cases where students chose a prompt action. (d) summarizes the prompting strategy codebook obtained during annotation. (e) shows the distribution of prompting strategies from the qualitatively coded prompt sample; cells are column-normalized fractions. We report both `first turn' (\textbf{n:first}, \textbf{i:first}) and `any turn' (\textbf{n:any}, \textbf{i:any}) bug-fixing turns; \textbf{n:first} is a subset of \textbf{n:any} and \textbf{i:first} is a subset of \textbf{i:any}.}
  \Description[RQ2 results on repair effort and prompting strategy by bug source]{Five-part mixed-method figure. Panels (a) and (b) are bar charts comparing natural versus injected bug groups for editing cases: mean edit distance and mean run-edit count are lower for injected-bug groups than natural-bug groups. Panel (c) is a bar chart of mean prompt length for prompting cases, with similar values across bug sources. Panel (d) is a two-column table defining eight prompting strategies, including task reframe, update signature, extend task, restate, provide guidance, provide output, provide code, and meta. Panel (e) is a heatmap showing the percentage distribution of these strategies across n:first, n:any, i:first, and i:any, with natural-bug groups emphasizing specification changes such as update signature and task reframe, and injected-bug groups emphasizing requests that directly guide the model or restate the task.}
  \label{fig.rq2}
\end{figure*}

To address our second RQ, we focus on how students repair buggy GenAI responses and how those behaviors differ by bug source (\textbf{n:first}, \textbf{n:any}, \textbf{i:first}, and \textbf{i:any}). Figure~\ref{fig.rq2} presents an overview of the results.

\textbf{Editing behavior.} \looseness-1 For edit actions, Figures~\ref{fig.rq2.a} and~\ref{fig.rq2.b} measure repair effort using mean absolute Levenshtein distance from the initial buggy output to the student’s final edited version, and the mean number of edit-and-run submissions within the turn. They show clear differences in editing effort. Following injected bugs, students make smaller code changes and fewer edit-and-runs than after natural bugs. In the `first turn' summaries, edit distance is lower for \textbf{i:first} than \textbf{n:first} (median $4$ vs.\ $24$, mean $21.96$ vs.\ $93.92$), and the same trend holds for `any turn' (median $4$ vs.\ $35$, mean $25.18$ vs.\ $121.02$ for \textbf{i:any} vs.\ \textbf{n:any}). Edit-and-run counts follow the same trend in means (first turn: $1.51$ vs.\ $2.15$; any turn: $1.55$ vs.\ $2.36$), with medians typically $1$ and small differences in some problems (e.g.,  $1$ vs.\ $2$ in Array IV/Binary for \textbf{i:first} vs.\ \textbf{n:first}, and in Matrix/Binary for \textbf{i:any} vs.\ \textbf{n:any}). Under the \textbf{n:first} vs.\ \textbf{i:first} conditions, per-problem Mann-Whitney tests indicate lower edit distance for injected than natural bugs in Array II, Array III, and Matrix (all $p<.001$), but not in Array IV ($p=.226$); Binary is interpreted descriptively due to limited data. For run-edit counts under the same conditions, per-problem tests yield $p=.030$ (Array III), $p=.049$ (Array IV), and $p<.001$ (Matrix), with $p=.061$ in Array II; Binary sparse as before.\footnote{Corresponding Mann-Whitney $U$ statistics (Array II, Array III, Array IV, Matrix) are: edit distance ($34990$, $50256.5$, $2218$, $2921.5$); run-edit counts ($22316.5$, $35118$, $2351$, $2626.5$). Binary too sparse.}

These results suggest that students treat injected bugs as localized repair tasks, possibly due to their confined nature. Students typically patch the small faults in near-miss solutions. In contrast, natural bugs often require broader code revisions or reworking, producing larger edits and potentially longer repair cycles.

\looseness-1\textbf{Prompting behavior.} Figure~\ref{fig.rq2.c} shows prompt length. It does not show systematic differences between bug source. In the pooled `first turn' summaries, prompt length is similar for \textbf{n:first} and \textbf{i:first} (median $138$ vs.\ $128$, mean $173.29$ vs.\ $177.25$), and the pooled `any turn' summaries are also close (median $149$ vs.\ $148$, mean $200.99$ vs.\ $192.47$ for \textbf{n:any} vs.\ \textbf{i:any}). Under the \textbf{n:first} vs.\ \textbf{i:first} conditions, per-problem Mann-Whitney tests also show mixed evidence: Array II ($p=.063$, medians $74$ vs.\ $87$), Array III ($p=.035$, medians $144$ vs.\ $102$), Array IV ($p=.422$, medians $180$ vs.\ $169$), and Matrix ($p=.034$, medians $154$ vs.\ $231$), with Binary too sparse\footnote{Corresponding Mann-Whitney $U$ statistics (Array II, Array III, Array IV, Matrix) are: prompt length $(9870,\,13542,\,2885.5,\,1105)$. Binary too sparse.}. We therefore focus on prompting strategy rather than prompt length.

\looseness-1Figure~\ref{fig.rq2.d} summarizes the reconciled prompting strategy codebook obtained during qualitative coding (Krippendorff's $\alpha=0.87$), while Figure~\ref{fig.rq2.e} shows the distribution of prompting strategies by bug source (\textbf{n:first}, \textbf{n:any}, \textbf{i:first}, \textbf{i:any}). Natural bug states are dominated by specification-oriented strategies, while injected bug states show more guidance and restatement-oriented strategies. In \textbf{n:first}, prompting strategy is led by \emph{Update signature} ($30.6\%$) and \emph{Extend task} ($27.8\%$). In \textbf{n:any}, the distribution shifts toward task-level clarification, with the majority being \emph{Task reframe} ($22.1\%$). In \textbf{i:first}, prompting strategies are more mixed, with \emph{Provide guidance}, \emph{Extend task}, and \emph{Restate} having similar frequencies. In \textbf{i:any}, \emph{Provide guidance} becomes the dominant strategy ($33.0\%$), while task reframing and signature updates are rare.

\begin{figure*}[t!]
\centering
\captionsetup[subfigure]{justification=centering,singlelinecheck=false}

\begin{subfigure}[t]{\linewidth}
    \centering
    \includegraphics[width=\linewidth, trim={0 0 0 0}, clip]{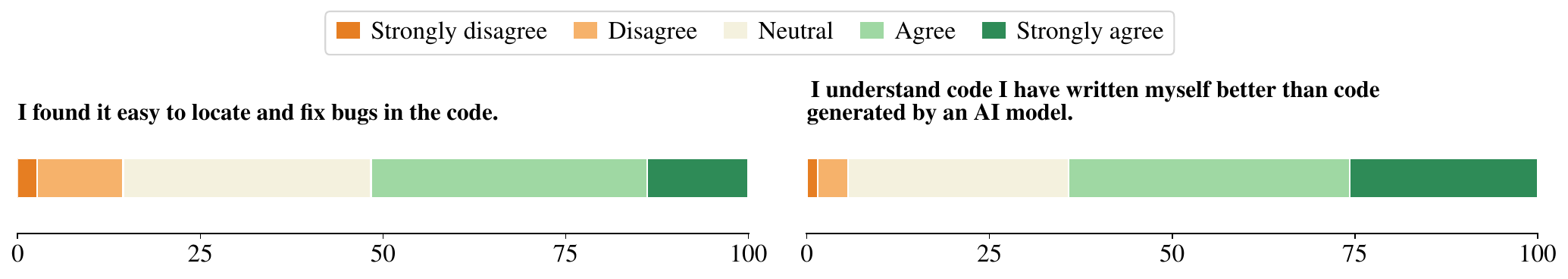}
    \vspace{-5mm}
    \caption{Likert responses for Q1 and Q2.}
    \label{fig.rq3.likert}
    \vspace{3.5mm}
\end{subfigure}
\begin{subfigure}{0.55\linewidth}
    \centering
        \scalebox{0.93}{
        \setlength\tabcolsep{2pt}
        \renewcommand{\arraystretch}{1.78}
        \begin{tabular}{@{} l >{\raggedright\arraybackslash}p{0.82\linewidth} @{}}
        \toprule
        \textbf{Qualitative code} & \textbf{Description} \\
        \midrule
        Code understanding & %
        Reading, tracing, and reviewing code to make sense of how it works. \\
        
        Debugging &
        Locating bugs and deciding what to try next after a failure. \\
        
        AI workflow strategy &
        How to collaborate with the AI and when to switch to editing or writing code directly. \\
        
        Specification & 
        Understanding requirements and communicating intent clearly. \\
        
        AI limitations &
        Limitations of AI-generated code and the need for critical evaluation. \\
        
        Solution exploration &
        Considering alternative approaches for solving the problem. \\
        
        Concept learning &
        Learning or strengthening programming concepts and precision. \\
        
        Bug types &
        Exposure to common or new bug patterns. \\
        
        Other &
        Vague or uncategorizable comments. \\
        \bottomrule
        \end{tabular}
        }
    \caption{Qualitative coding scheme used for Q3.}
    \label{fig.rq3.codebook}
\end{subfigure}
\hfill
\begin{subfigure}{0.38\linewidth}
    \centering
    \includegraphics[width=\linewidth]{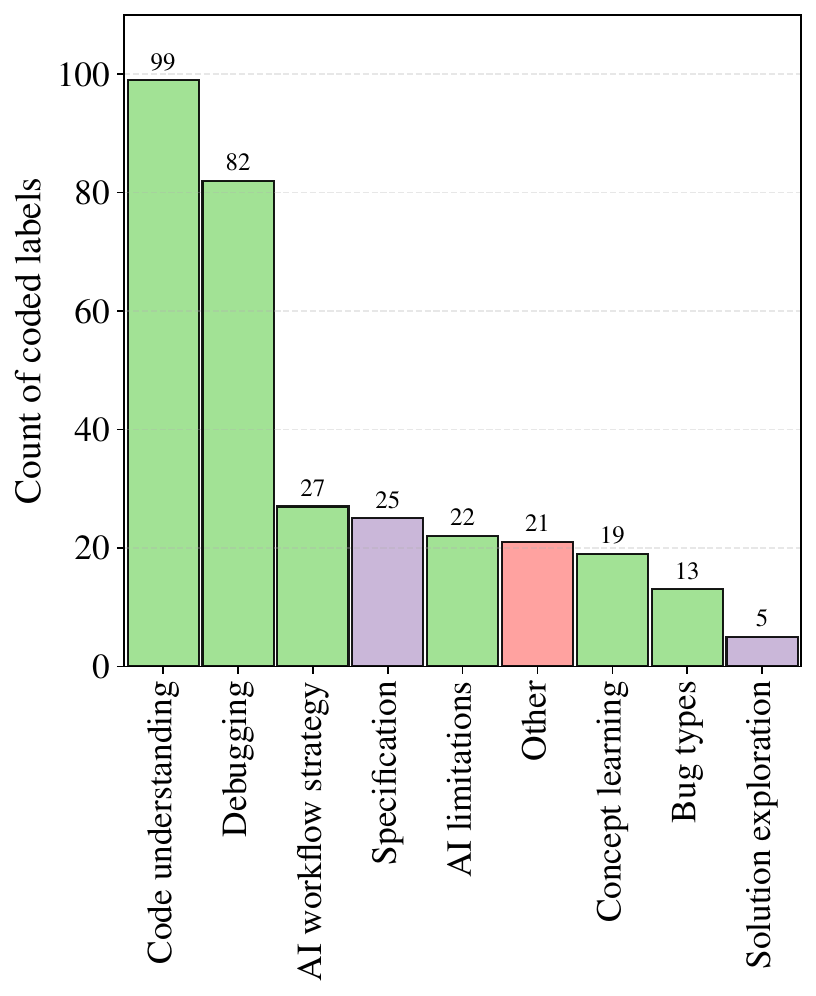}
    \caption{Qualitative code frequencies for Q3.}
    \label{fig.rq3.themes}
\end{subfigure}

\caption{
\looseness-1RQ3 reflections on learning with buggy GenAI code. (a) shows perceived ease of locating and fixing bugs and comparative understanding of self-written vs. AI-generated code (Q1–Q2). (b) summarizes the reflection codebook. (c) reports perceived learning benefits from open-ended reflections (Q3) as code frequencies, with green for bug injection and purple for general prompt-based programming.
}
\Description[Reflection results: Likert distributions and coded learning themes]{Three-part figure summarizing student reflections. Panel (a) shows two horizontal stacked bar charts for Likert responses. One corresponds to perceived ease of locating and fixing bugs, and the other to understanding self-written code better than AI-generated code; both show responses spread across the five Likert categories, with stronger agreement on the self-written-versus-AI understanding item. Panel (b) is a table listing qualitative codes and short definitions, including code understanding, debugging, AI workflow strategy, specification, AI limitations, solution exploration, concept learning, bug types, and other. Panel (c) is a bar chart of code frequencies, with the largest counts for code understanding and debugging, followed by AI workflow strategy and specification.}
\label{fig.rq3}

\end{figure*}

\looseness-1This contrast suggests differences in how students respond to the two contexts. Natural bugs appear to signal a misunderstanding of the task, possibly encouraging students to revise or clarify their task description. Injected bugs seem to signal a near-miss solution, possibly encouraging students towards more guidance-oriented requests to the GenAI assistant.

\looseness-1\textbf{RQ2 summary.} Overall, our results highlight differences between bug sources in how students proceed after a failure, beyond final outcome. We more often see localized repair and guidance-oriented prompting after injected bugs, whereas natural bugs more often come with specification-focused revisions and task-level clarification prompting strategies.

\subsection{RQ3: Student Reflections on Debugging, Code Understanding, and Learning Benefits}

In our third RQ, we complement the analysis of students' behavior with their perceptions. We analyze student reflections collected after task completion, Figure~\ref{fig.rq3} presenting the results. Specifically, Figure~\ref{fig.rq3.likert} shows their responses to the first two Likert-scale questions, Q1 and Q2. For Q1, many students selected neutral ($\sim34\%$), but responses still leaned slightly positive overall. Just over half of students agreed or strongly agreed ($\sim52\%$) that they found it easy to locate and fix bugs, suggesting that this type of bug-fixing exercise was generally manageable for students. For Q2, agreement was stronger. Roughly two thirds agreed or strongly agreed ($\sim64\%$) that they understand self-written code better than AI-generated code, with very little disagreement ($< 6\%$). This perceived gap between self-written and AI-generated code is also emphasized in many of the open-ended reflections, which we discuss next.

\looseness-1To analyze responses to the free-text question, Q3, we used thematic coding. We first  conducted open coding to identify recurring themes and then consolidated these into a closed coding scheme. Figure~\ref{fig.rq3.codebook} summarizes the reconciled theme set, and Figure~\ref{fig.rq3.themes} reports label frequencies after disagreement reconciliation. Agreement was strong overall (mean Krippendorff's $\alpha=0.82$, MASI-based $\alpha=0.80$), but was lower for \emph{Concept learning} ($\alpha=0.61$) and \emph{Bug types} ($\alpha=0.59$), which were less frequent and harder to distinguish consistently, so we interpret those categories more cautiously.

\textbf{Debugging and code comprehension.} \looseness-1Many responses emphasized \emph{Code understanding} and \emph{Debugging} as primary learning benefits. Students often framed the activity as learning by diagnosing and recovering from failure as opposed to simply producing a solution. One student described it as \textit{``[...] learn[ing] by failing [...] you get a close enough result but you learn the process of debugging which is a vital skill to have, especially with team projects''}. Another highlighted the value of careful tracing and inspection, noting that it \textit{``helps with explicitly `visualising' code (i.e. executing it in my head) and helps with error-spotting, in the same sense that training chess puzzles is beneficial practice for finding tactics in real chess games [...]''}. Similarly, a student reflected, \textit{``I think it forces you to actually really understand the nuances in the syntax because the tiniest errors will be what throw you off. Being able to see these errors quickly and get good at finding them is very useful. It also allows you to improve your debugging skills with more generic bugs rather than just the ones you are always making yourself. That way you learn how more bugs arise, and can become a more efficient coder who could also read other people's code and fix it''}. Overall, students reported practicing reading, tracing, and repairing plausible programs, suggesting the activity supported verification-centered skills alongside prompt refinement.

\textbf{AI workflow and its drawbacks.} \looseness-1Students also reflected on \emph{AI workflow strategy} and \emph{AI limitations}, often emphasizing that effective use of the assistant requires active oversight and selective adoption of suggestions. One student described settling on a cautious workflow, writing: \textit{``I realised that the best way to write code is likely to try it yourself for the first time, then if you hit a roadblock consult the AI but only implement code that you understand, because if you do not then it is very easy to lose track of the code and mess it up. I found it is harder to interpret finished code then a code you wrote yourself of course''}. Others reflected on unreliability and the need for human oversight. For example, one student (after being reminded that bugs are injected on purpose) wrote: \textit{``Really... it's a model designed to intentionally make mistakes... I had my suspicions. I think that's a bit of an issue a lot of different LLM's are facing currently. The root of the issue I believe is that because of the fact that now LLM's are being trained by each other, they often pass on bad habits such as the fact that if an output is deliberately wrong, it incentivises the user to submit another prompt [...] (more user engagement). [...] Because of the fact that this is such a prevalent issue, it is good to learn how to work around it''}. Another student summarized this dependency succinctly: \textit{``That AI requires human assistance to make good code''}. Overall, students viewed AI code as requiring human oversight, and the activity helped them decide when to reprompt, inspect, or repair locally.

\textbf{Specification and broader learning.} \looseness-1Beyond the above themes, several reflections emphasized \emph{Specification}, \emph{Concept learning}, and exposure to \emph{Bug types}. Some students noted that the activity pushed them to be more explicit about intended behavior and its implementation, for example: \textit{``It helps me understanding specifically on what I'm writing down and what my code should look like''}. Others described practicing core programming skills and building a better grasp of how things work, noting: \textit{``It was better because I felt like I had to use my own skills more, and it gave me more practice at figuring out how certain things worked''}. Finally, students also valued encountering a diversified range of mistakes, as one student put it: \textit{``Gives me opportunity to fix coding errors that i would maybe not encounter coding by myself''}. Overall, the activity reinforced links between specification, implementation, and failure analysis, beyond debugging alone.

\textbf{RQ3 summary.} Our results show that students experienced the activity as practice in code understanding, debugging, and AI workflow strategy, and not just as a harder way to complete tasks. In the Likert responses and open-ended reflections, students emphasized that AI-generated code was harder to understand than self-written code and still required review and debugging. Dominant themes from the open-ended reflections support the instructional value of deliberate, runnable bugs in prompt-based programming activities.

\section{Discussion}
\label{sec:discussion}

Our original goal with this research was to explore whether deliberately injecting bugs into a GenAI workflow would encourage students to engage in code review. Our results indicate that injected bugs are more likely to be resolved immediately than natural bugs, with immediate success rates much higher following injected bugs (e.g., $90\%$ vs.\ $51\%$ on first turns). This is consistent with the fact that injected bugs were introduced only after the assistant first produced correct code, so they were usually closer to a working solution and left a smaller repair to the student. In that sense, the observed behavioral difference is understandable. Students seem more willing to edit when the code is already near correct, whereas natural bugs more often coincide with specification mismatches that are more naturally addressed through reprompting. At the same time, this is precisely the design space we aimed to create, one in which students must inspect, verify, and locally repair generated code instead of relying only on repeated prompting.

\looseness-1We now interpret students' follow-up actions through two design-relevant lenses grounded in their behavior and reflections. Prompt-based programming lets students iterate on natural-language specifications and patch code artifacts in the same workflow \cite{DBLP:conf/sigcse/00010PLABR24,DBLP:conf/kolicalling/ReevesP00MLNB25,DBLP:conf/iticse/VadapartyZSPAB024,padurean2025prompt}. This flexibility, although powerful, changes the meaning of `debug', because students must decide whether to refine a specification, localize and repair a defect, or verify a plausible-looking solution before trusting it \cite{DBLP:journals/cacm/DennyPBFHLLRSS24,DBLP:journals/corr/abs-2402-01580}. Across RQ1–RQ2, bug source is associated with where students direct repair effort and how they follow up via prompts and edits. From RQ3, injected near-misses create opportunities for verification, which students often describe as useful practice in code understanding and debugging. We use these patterns to motivate two implications.

\subsection{Diagnosis Before Action: Bug Source Shapes the Scope of Repair}

Our RQ1 findings show that students do not respond to buggy GenAI code with a single default action. Instead, follow-up actions shift with bug source, suggesting that students may use the assistant's response as one cue for the scope of repair. RQ2 strengthens this interpretation by showing differences in kind, not just amount.  Some prompts clarify specifications or constraints, while others focus on diagnosing failures or guiding the assistant toward a fix. Code edits show a similar split, ranging from small, localized patches to larger revisions and longer repair loops. One interpretation is that students use buggy outputs as signals about what needs repair -- the task description or the code artifact -- and they take their next action accordingly. In GenAI programming, where both options remain available, better diagnosis may help students avoid unproductive loops that apply the wrong kind of repair.

\looseness-1A useful way to frame this behavior is as a metacognitive control decision under uncertainty. Learners monitor evidence such as test results, console output, and the alignment between the code and the specification, and they decide whether a failure reflects a specification mismatch or a local implementation defect. According to self-regulated learning theory, such monitoring and interpretation of evidence supports decisions about how to proceed when outcomes diverge from expectations \cite{WinneHadwin1998SRL,zimmerman2002srl}. Metacognitive experience models similarly emphasize that felt difficulty and confidence can influence action selection under uncertainty, rather than implying one fixed response pattern \cite{efklides2011masrl}.  This distinction aligns with prompt-versus-edit choices in GenAI-assisted programming: prompting revises the specification the model uses to generate code, while editing targets a concrete artifact that can be inspected and repaired locally \cite{mccauley2008debugging,DBLP:conf/icse/KoM08}. From this perspective, the pedagogical target is not prompting or editing as isolated skills, but diagnosis that justifies the next step.

\looseness-1Our findings suggest that students interpret bug sources and proximity to a correct solution as signals about how to proceed, which has implications for interface design. One practical scaffold is to ask students to commit to an intent, i.e., specification repair or code repair, and provide a checkable artifact that matches the intent before the system enables the next action. Help design research shows that learners often need support to seek the right kind of help at the right time, and that interface-level nudges can shift help use from unproductive dependency toward useful progress \cite{DBLP:conf/sigcse/Edwards04,Aleven2003HelpDesign,NelsonLeGall1981HelpSeeking}. In GenAI workflows, lightweight feedback that connects observed failures to explicit hypotheses may encourage deliberate inspection and reduce blind trial-and-error \cite{DBLP:conf/sigcse/YehTGYFC25,DBLP:conf/icse/KoM08}. Debugging intervention work similarly suggests that teaching a systematic process can improve debugging performance and reduce unproductive struggle, which makes diagnose-before-action a natural approach to teach alongside prompting skills \cite{DBLP:conf/wipsce/MichaeliR19,DBLP:journals/toce/YangBOBS24,Reiser2004Scaffolding}. A bug-source-aware design could suggest default repairs while allowing these to be overriden, maintaining autonomy while nudging learners toward likely productive moves \cite{WinneHadwin1998SRL,DBLP:conf/sigcse/YehTGYFC25}. More broadly, prompt-centered assignments should assess and scaffold diagnostic reasoning, not only prompt wording, because recovery success depends on acting on the right object, at the right time, with the right evidence \cite{DBLP:conf/sigcse/00010PLABR24,DBLP:conf/icer/YangZXBS24,mccauley2008debugging}.

\subsection{Designed Near-Misses as Verification Scaffolds in GenAI Workflows}

Our RQ1–RQ2 findings show that injected, runnable bugs do more than introduce errors, they change how students respond. Near-miss outputs are close enough to a correct solution that they invite inspection and targeted repair rather than full regeneration via prompting. In our data, injected bugs are linked to different action choices and response patterns than natural bugs, and edits after injected bugs are typically smaller and more localized. A plausible explanation is that near-misses make local repair both feasible and visible, signaling that the productive move is to verify and patch the code rather than continue prompting. RQ3 reflections reinforce this interpretation, as students describe the exercise as practice in reading code, tracing behavior, and debugging under realistic constraints. These results support a design claim: near-miss injection turns a generative workflow into a continuous practice environment for code verification, embedding diagnostic repair into the core interaction rather than treating it as a separate task. This is important because AI-generated code can look plausible while still being wrong, and verification remains a persistent challenge \cite{DBLP:conf/chi/Vaithilingam0G22,sarkar2022what,DBLP:journals/infsof/OertelKH25}. Large-scale studies of GenAI-generated code also report non-trivial correctness and bug prevalence, making verification a core skill rather than an optional add-on \cite{tambon2025bugs,mo2025assessing,DBLP:journals/ese/TihanyiBFJC25}.

One interpretation of these patterns is that near-miss solutions create a form of manageable challenge, as they require students to critically read the code while still keeping the repair space manageable. Research on error management and productive failure suggests that there are benefits to learning when learners engage constructively with bounded difficulty~\cite{keith2008effectiveness,steele2014error,Kapur2008ProductiveFailure}. Prior work on desirable difficulties similarly argues that learning tasks should be challenging enough to trigger deeper processing, but not so challenging that they lead to unproductive struggle~\cite{Bjork1994Metamemory,bjork2011making}. In programming education, effective debugging involves forming hypotheses, testing them, and revising based on evidence rather than making superficial edits~\cite{mccauley2008debugging,DBLP:journals/hhci/KatzA87}. Formative feedback research adds that effective feedback should be specific and actionable~\cite{shute2008focus,HattieTimperley2007PowerFeedback}.  In our setting, runnable near-misses paired with immediate test feedback can support this bounded verification loop, where learners can form a local hypothesis, test it, and iterate within the same workflow, aligning the activity with evidence-driven debugging rather than trial-and-error~\cite{DBLP:conf/icse/KoM08,DBLP:conf/sigcse/Edwards04}.

\looseness-1Design implications follow from treating bug injection as a verification scaffold that should be clear, low-friction, and fadeable \cite{Reiser2004Scaffolding}. One practical design is to make verification mode explicit when a near-miss appears, and require a minimal piece of evidence before enabling full regeneration. For example, the system could ask for one checkable artifact, such as a failing test case, an expected property, or a short hypothesis about a suspect location, before allowing another GenAI solution. This keeps help-seeking instrumental and ties assistance to evidence, which aligns with help-seeking theory and help-design guidance \cite{NelsonLeGall1981HelpSeeking,Aleven2003HelpDesign}. Recent classroom systems similarly suggest that structured interactivity can help learners work productively with LLMs by encouraging testing, inspection, and response to concrete signals instead of treating the model as an oracle \cite{DBLP:conf/sigcse/YehTGYFC25,DBLP:conf/sigcse/RenzellaVST25}. At the same time, reflection should be used carefully. Reflection can improve monitoring and evaluation, but it can also add friction and reduce satisfaction \cite{DBLP:journals/corr/abs-2512-04630}. Cognitive load theory suggests that overloaded learners may fall back on fluency-based heuristics, including trusting outputs that only \emph{look} correct \cite{sweller1988cognitive}. Self-explanation cues can counter shallow processing by requiring learners to articulate why code should work, but these cues need to stay lightweight to avoid becoming another source of friction \cite{chi1994eliciting}. A practical balance is to pair near-miss injection with lightweight verification cues that are easy to complete and fade, while making deeper reflection optional or triggered by repeated difficulty \cite{shute2008focus}. This preserves an authentic workflow while turning ``verify before you trust'' into a practiced habit \cite{DBLP:journals/corr/abs-2402-01580,DBLP:conf/chi/Vaithilingam0G22,DBLP:journals/infsof/OertelKH25}.

\subsection{Limitations}

\looseness-1We acknowledge our findings should be interpreted in light of several limitations and give ideas for clear next steps. First, the study is tied to a specific setting, an introductory C course and a set of five platform problems, out of which only two were required to be solved for credit. Future work should aim to replicate the findings across languages, institutions, and more open-ended tasks to test generalizability. Second, follow-up actions are shaped by the platform design and the kind of feedback students get in a prompt-and-test workflow with hidden tests. Future work should vary the feedback and tools available. For example, it could explain a small counterexample, provide more detailed run-time feedback, let students write their own tests, or add lightweight debugging support, and then check whether the same bug source patterns still hold. Third, the current study did not disclose the source of bugs to students. Future work should examine how different levels of transparency regarding bug injection affect student trust and engagement. Finally, because this was an in-class deployment rather than a controlled comparison between conditions, differences we observe may also reflect course context and platform choices, not only bug source. A natural next step is an A/B study that changes one factor at a time, such as injected bugs on versus off, different injection rates, or editing enabled versus disabled, and evaluates both in-session behavior and later outcomes such as performance on new problems or delayed follow-up tasks. Another important direction is to test whether these workflows lead to measurable learning benefits, for example through controlled pre-test/post-test designs that assess code comprehension, debugging performance, and transfer to new tasks.

\section{Conclusion}
\label{sec:conclusion}

\looseness-1In this paper, we studied how introductory programming students respond to bugs while solving prompt-based programming tasks with a GenAI assistant. Across $2{,}636$ analyzed sessions and $6{,}071$ turns where a bug needed to be fixed, we found that students' next actions are strongly shaped by the type of bug they encounter. Natural bugs are more often associated with specification-focused follow-up work, while injected near-miss bugs are more often associated with inspecting and repairing code. This difference is also visible in prompt content, which shifts between clarifying task intent and constraints versus providing more detailed guidance for a targeted fix. Students' reflections further suggest that combining prompting with structured opportunities for review and repair can support verification of GenAI outputs and awareness of GenAI limitations. More broadly, our results suggest that mixing natural failures with realistic runnable near-misses can support both specification refinement and careful code review and debugging within the same workflow. Looking ahead, these findings motivate prompt-based programming activities and platforms that scaffold diagnosis and verification alongside prompting. One direction is to add lightweight cues that help learners clarify the task after repeated mismatch signals. Another is to trigger debugging-oriented checks when localized defects are more likely, for example asking for a brief hypothesis, a suspect location, or a small validating test. Overall, our findings suggest that adding runnable near-misses can make verification and repair a more consistent part of prompt-based programming practice.

\begin{acks}
This work was supported by Research Council of Finland grant \#356114. Funded/Cofunded by the European Union (ERC, TOPS, 101039090). Views and opinions expressed are however those of the author(s) only and do not necessarily reflect those of the European Union or the European Research Council. Neither the European Union nor the granting authority can be held responsible for them.
\end{acks}

\bibliographystyle{ACM-Reference-Format}
\balance
\bibliography{main}

\end{document}